%% file: 0_main.tex
\documentclass[nonacm,sigconf]{acmart}

\usepackage[utf8]{inputenc}%(only for the pdftex engine)
\usepackage{amsmath}

\usepackage{filecontents}

\usepackage[]{xcolor}
\PassOptionsToPackage{hyphens}{url}\usepackage{hyperref,wasysym}

\usepackage{multirow}
\usepackage[export]{adjustbox}
\usepackage{tabularx}

\usepackage{color, colortbl}
\definecolor{Gray}{gray}{0.90}

\usepackage{booktabs}

\usepackage{csquotes}

\usepackage{subfigure}
\usepackage{balance}
\usepackage{enumitem}
\setcopyright{none}
\renewcommand\footnotetextcopyrightpermission[1]{} % removes footnote with conference information in first column
\begin{document}

\title{{`Surprised, Shocked, Worried’}: User Reactions to Facebook Data Collection from Third Parties}

\author{Patricia Arias-Cabarcos}
%\orcid{0000-0001-7401-6185}
\affiliation{%
  \institution{Paderborn University}
  \country{Germany}
  }
  \email{pac@mail.upb.de}
  \author{Saina Khalili}
\authornote{Contributed to this research while doing her Master Thesis at KIT in 2019 (Currently unaffiliated).}
\affiliation{%
  \institution{KIT} \country{Germany}
  }
  \email{saina.khalili72@gmail.com}
\author{Thorsten Strufe}
\affiliation{%
  \institution{KASTEL Security Research Labs, KIT}\country{Germany}
    }
\email{strufe@kit.edu}

\begin{abstract}
Data collection and aggregation by online services happens to an extent that is often beyond awareness and comprehension of its users. Transparency tools become crucial to inform people, though it is unclear how well they work. To investigate this matter, we conducted a user study focusing 
    on Facebook, which has recently released the ``Off-Facebook Activity'' transparency dashboard that informs about personal data collection from third parties. We exposed a group of n = 100 participants to the dashboard and surveyed their {level of awareness} and reactions to understand how transparency impacts users' privacy attitudes and  {intended} behavior. Our participants were surprised about the massive amount of collected data, became significantly less comfortable with data collection, and more likely to take protective measures.  Collaterally, we observed that current consent schemes are
    inadequate. Based on the survey findings, we make recommendations for more usable transparency and highlight the need to raise awareness about transparency tools and to provide easily actionable privacy controls.
\end{abstract}
%MAJOR POTENTIAL
 %Based on the survey findings  {and additional heuristic-based usability evaluations}

\keywords{usable privacy, social networks, transparency, user study}

\maketitle  

\section{Introduction}
\input{introduction}

\section{Background and Related Work}
\subsection{Off-Facebook Activity}

\input {OFA}

\input{RW}

\section{Methodology}
\input{Methodology}

\section{Results}
\input{Results}

\section{Discussion}
\input{Discussion}

\section{Conclusion}
\input{Conclusion}

\begin{acks}

The authors gratefully acknowledge funding by the Helmholtz Association
(HGF) through the Competence Center for Applied
Security Technology (KASTEL), topic “46.23 Engineering
Secure Systems”. We thank the anonymous reviewers who helped to improve this paper with their
useful feedback.
\end{acks}

\bibliographystyle{ACM-Reference-Format}
\bibliography{biblio}

\appendix

\section{Survey Instrument}
\input{Survey}

\end{document}

%% file: introduction.tex
Online social networks collect and process a great deal of personal information, mainly to target users with personalized content and advertisements, making profit from their interactions. The abuse of these data can have a direct impact on user autonomy and agency~\cite{andre2018consumer,ward2018social}, and even disrupt society in unprecedented ways, as revealed by scandals like the Cambridge Analytica interference with the US elections in 2016~\cite{cadwalladr2018revealed}. Facebook (FB) in particular, the most used social media platform, has created a vast tracking infrastructure that silently collects user data as they navigate the web and use applications, through ubiquitous\footnote{ In April 2018  around 8.4 million websites were using the Like button, another million the Share button, and 2.2 million websites had Facebook Pixels~\cite{FBtracking}} social plugins (e.g., ``Like'' buttons), hidden pixels, and other business tools offered to third party companies~\footnote{\url{https://www.facebook.com/help/331509497253087}}, which extend to collecting information about offline activities, like sales in brick and mortar shops\footnote{\url{https://www.facebook.com/business/learn/facebook-offline-conversions}}. %But also from offline user activities~\ref{ofline}. 
 Recent research \cite{cabanas2020does} uncovered that, based on these data, Facebook assigned ad preference labels that correspond to `special categories' of sensitive data to 73\% of EU FB users, even after the GDPR \cite{GDPR} was enacted. Sensitive categories were also attributed to users in countries where disclosure of certain information (e.g., sexual orientation) can be life threatening. %tag
 %They estimate the cost of unveiling the identity of FB users labeled with potentially sensitive ad preferences may be as cheap as e0.015 per user. FDTV
  This form of aggressive attribution for targeting purposes poses important threats to privacy, and it fosters discrimination based on race, gender, or age, among other sensitive attributes~\cite{angwin2016facebook, speicher2018potential,andreou2018investigating}.
%statista2 
%https://www.statista.com/statistics/272014/global-social-networks-ranked-by-number-of-users/
%even if no interaction with buttons
%Targeted advertising poses important threats to privacy as it can enable discrimination based on sexual orientation, ethnic origin, gender among other sensitive groups [5, 20, 39, 87]. Against this massive surveillance, the orivacy literature says, users
%Do users care?
%\textcolor{red}{need more refs} 

Amidst this surveillance reality, users do little to protect themselves~\cite{young2013privacy, schaub2016watching}. The privacy literature points out that social network users seem to care more about \textit{social privacy} (the concern about controlling access to personal data by other people), than about \textit{institutional privacy} (the concern about how providers and third parties will use personal data) \cite{young2013privacy}. However, user inaction against institutional surveillance likely stems from a poor understanding about how tracking works~\cite{andreou2018investigating, dolin2018unpacking, schaub2016watching, nurse2017behind}. Transparency is therefore crucial to inform and empower users, but this type of feature is rarely available within social networks or just provided in a limited manner. It was not until January 2020 that FB released the ``Off-Facebook Activity'' transparency tool, a dashboard for users to see which third party apps and websites have shared their information with Facebook to expand the aggregated digital dossiers that represent the user profiles~\cite{fowler2020facebook}. Given the knowledge gap on how FB transparency can affect users behavior and explain their (lack of) privacy-protective strategies, we decided to explore the following research questions:

\begin{itemize}[leftmargin=*]
\itemsep0.15em 
    \item[] \textbf{RQ1} \textbf{[Awareness and Impact on Privacy Attitudes]} How does the ``Off-Facebook activity'' tool affect users' privacy attitudes?
    \item[] \textbf{RQ2} \textbf{[User Reactions and Intended  Behavior]} What are users' reactions and  {intended} behavioral changes after exposure to the ``Off-Facebook activity'' tool?
    
    \item[] \textbf{RQ3} \textbf{[Usability and Usefulness]} How usable is the ``Off-Facebook activity'' dashboard as a transparency enhancing tool as evaluated by users? 
\end{itemize}

For the first two questions, we hypothesize that more knowledge, gained through the transparency tool, increases privacy concerns and leads to privacy-protecting behavior, based on the ``bounded rationality" explanation for the privacy-paradox \cite{barth2017privacy}. The third question is motivated by the apparent complexity of the FB transparency tool --plagued with dark patterns and complex menus-- since low usability is a recognized impediment to efficiently use security and privacy tools~\cite{sasse2001transforming, garg2014privacy}. Therefore, we seek to understand how to improve usability for more efficient transparency.

 To address our research questions, we conducted an online survey with N = 100 participants. Users were first asked about their privacy attitudes and protective strategies, then directed to the ``Off-Facebook Activity'' dashboard to visualize and navigate the data collected by FB about them. After exposure to the tool, we asked them again about privacy attitudes, including questions on their reactions, intended behavior and how usable the tool was to achieve their privacy goals.    
  
  Our findings show that users want transparency tools and that they would like these tools to be not only easy to use, but also easy to access and more actively advertised to the public. 
  Despite most users where active in FB, 85\% of the participants did not know about the ``Off-Facebook Activity'' tool more than a year after its release.
 Once participants are exposed to transparency information, their self-reported likelihood to take protective measures against FB tracking increases. They met the amount of collected data with shock and surprise, which makes it questionable if the consent given to share the data can be considered informed, and hence effective. On the positive side, our participants  experienced positive feelings related to the possibility of exercising control, though  there are limitations that hinder user understanding and ability to take action. Based on these findings,
  we contribute to the literature with recommendations on how to improve the design of transparency tools.

%% file: OFA.tex
\label{sec:OFA}

\begin{figure*}[ht!]
   
    \includegraphics[width=1\textwidth]{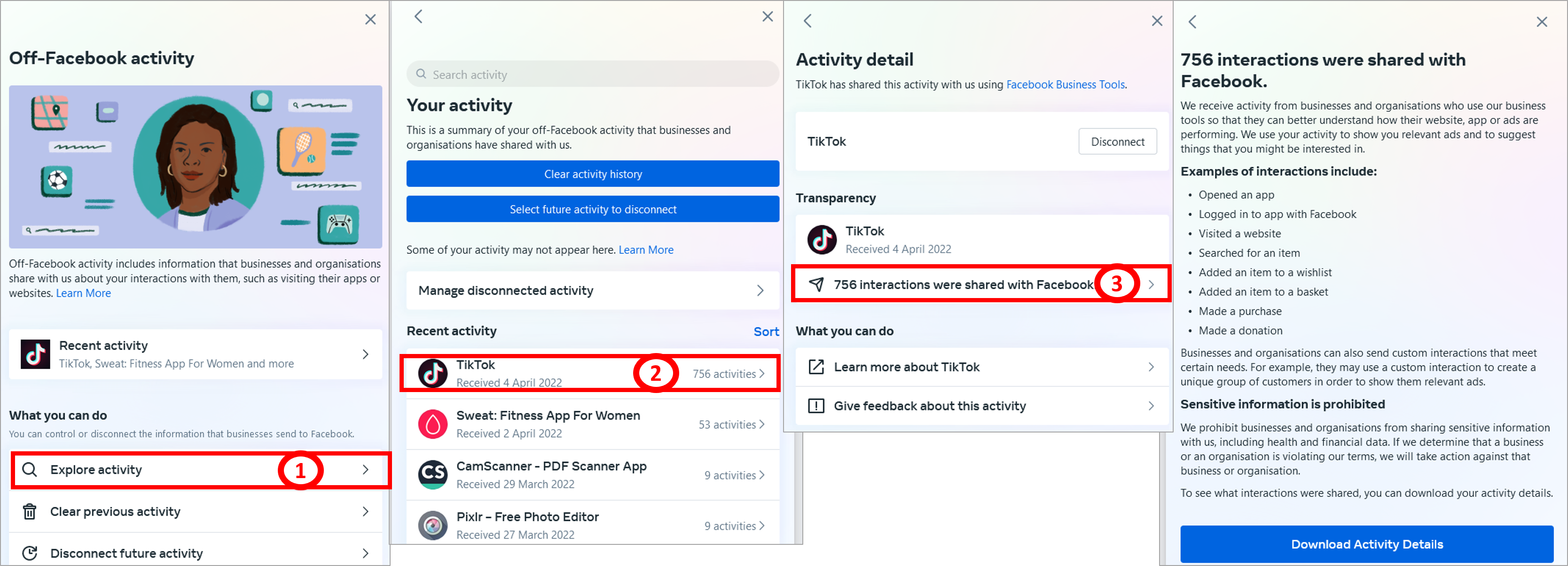}
        \caption{Interface of the Off Facebook Activity transparency tools. It shows screenshots of the sequence of steps to visualize the list of 3rd parties that share user data, and access detailed information about the activity with a specific app.  \label{fig:interfaces} }
  \end{figure*}

 {
Facebook rolled out the \textit{``Off-Facebook Activity''} tool in August 2019, initially in a few countries, until it became globally available in January 2020\footnote{\url{https://about.fb.com/news/2020/01/data-privacy-day-2020/}}. The goal of the tool is to provide transparency and control regarding the data that 3rd parties send to Facebook about their users. The tool's landing page (see Fig. \ref{fig:interfaces}) gives access to a summary of the apps and websites that share data with Facebook, where the user can see the number of interactions involved, and a generic list of the type of data that might have been exchanged. To obtain details about the specific information sent, the tool links to a data export page where users can download JSON or HTML files to navigate offline. As mechanisms for control, \textit{``Off Facebook Activity''} lets users clear their previous activity and disconnect their future activity, options that eliminate the link between the data and the user account, only for past exchanges or for any past and upcoming exchange, respectively.} Our research on this tool extends the body of knowledge on privacy in Social Networks~\cite{zheleva2011privacy, paul12c4ps, kayes2017privacy}, building on and complementing prior work on tracking transparency.

%% file: RW.tex
\label{sec:RW}

\subsection{Privacy in Social Networks}

With the development and continued increase of social media usage, privacy concerns have been on the rise and a highly debated and investigated topic. Many researchers posit that there is a ``privacy paradox''~\cite{barnes2006privacy} in that users expose their personal information in these platforms even if they claim to worry about privacy. 
 
 Prior work in this area explored different factors shaping the apparent paradox. A survey study on the privacy strategies of Facebook users with 384 participants~\cite{garg2014privacy} showed that behavior is biased by the lack of information, as people do not know enough to grasp privacy risks and so they have to make ``bounded rational decisions''. Another key factor observed was that the usability of privacy controls is a deterrent for FB users to apply privacy measures. 
 
 A lion share of this literature has focused on the social dimension of privacy (control of personal information during social interactions) instead on the institutional side of it (control of information usage by institutions/3rd parties). Indeed, despite the large body of research and development projects on protecting users from the providers \cite{gurses2013two, jahid11easier, cutillo10security, guidi18managing}, privacy enhanced alternatives have not managed to gain much traction, so far. In this regard, Young et al.~\cite{young2013privacy} observed that social site users seem to be less concerned with institutional privacy because they report little protective actions against surveillance. But this work also suggests that users might not be aware of the applied surveillance practices, given the limited institutional transparency in place. A later study on user privacy strategies~\cite{choon2018revisiting} confirms that social network users have but vague knowledge of institutional surveillance practices. We aim at extending this body of research by investigating users' awareness about institutional transparency, and the effect of transparency on privacy attitudes and behavior.  {We selected Facebook because of its popularity as the most used social network ($\sim$3 billion users\footnote{https://www.statista.com/statistics/264810/number-of-monthly-active-facebook-users-worldwide/}) and the consequential strong impact that transparency or the lack of it can have. The ``Off-Facebook Activity'' tool is unique in that it shows data sent to the social network by 3rd parties, while other transparency interfaces only agglutinate user activity within a single application or domain of applications (e.g., Google’s My Activity shows actions in Youtube, Gmail, etc.). We take a first step towards understanding reactions to this less known data sharing practices by social networks providers.}

\subsection{Transparency Tools}
%TETs survey
Transparency is an important privacy principle at the heart of informational self-determination. Users have the right to know how their data are collected and used, which is especially relevant given that the business model ingrained in the Web is based on user profiling. Transparency Enhancing Tools (TETs) provide technological means to this end~\cite{TETtools}. With regard to web tracking, the most common transparency tools  provide information about: 1) which trackers are present in a website, (e.g., ad blockers like Ghostery~\cite{Ghostery}), and 2) why specific ads are shown to a user (e.g., the AdChoices icon~\cite{komanduri2011adchoices}). Furthermore, there is extensive work on privacy or transparency dashboards to allow online service users exploring and managing data collected about them \cite{TETtools, janic2013transparency, raschke2017designing, schufrin2020visualization, urban2019your}.
 Research so far has shown that these tools are limited, often incomplete or imprecise, and sometimes misleading~\cite{weinshel2019oh, andreou2018investigating}. Weinshel et al.~\cite{weinshel2019oh} conducted a longitudinal field study with 425 users concluding that web tracking transparency tools with more detailed information (potential data inferences instead of just tracker presence) lead to increased intention to take privacy-protecting actions.  {We build on this study with the aim to extend and complement research knowledge in a different domain: the domain of social network transparency, specially focusing on 3rd party data collection.}

In the social network domain, Wei et al. \cite{wei2020twitter} explored user's perceptions about Twitter ad targeting mechanisms, finding out that they consider these practices invasive and prefer detailed explanations about how their data are used.  { Similar to this work, we see users of the Off Facebook Activity tool demand increased detail about 3rd party data collection from FB.Farke et. al~\cite{farke2021privacy} conducted a user study with Google My Activity, providing evidence that users concerns about Google’s data collection significantly decreased, and perceived benefit increased post exposure to My Activity. They also observed that most users would not use the features of the dashboard nor change their behavior. In contrast to these findings, we observe stronger discomfort with data collection after using FB's transparency tool, as well as increased intention to change behavior, which might be explained by the type of information shown (3rd parties outside FB). In fact, users are not only surprised by the amount and detail of collected data, as in the My Activity study~\cite{farke2021privacy}, but also about the number and type of sharers (e.g., well-reputed banks). }
 
 Furthermore, transparency tools and studies related to FB are scarce. As external tool,  González et al.~\cite{gonzalez2017fdvt} proposed a browser plugin to inform users about the real-time revenue they generate for Facebook while using it. Their Facebook Data Valuation Tool (FDTV) was efficient in raising awareness about the monetary value associated to user data. Subsequent extensions and studies with the same tool, uncovered that FB assigns sensitive ad preference labels (e.g., related to sexual orientation) to 67\% of users worldwide in 2019~\cite{cabanas2020does} and estimated the cost to identify user labels in only 0.015\euro ~\cite{cabanas2018unveiling}.   {These tools could be integrated and used together with FB native tools to provide a broader transparency picture.}

 We complement and go beyond the current literature that provides insights on user reactions  {to social network's transparency} by: 1) analyzing the impact of a different type of transparency (data collection from 3rd parties) on user attitudes and behavior, and 2) studying the role of transparency usability,  {which has not been yet explored in this specific domain.}

%% file: Methodology.tex
\label{sec:methodology}
This section explains our methodology to explore the impact of transparency about Facebook data collection on users' privacy attitudes and behavior,  {as well as to analyze the usability of the tool to satisfy users' transparency goals}. We include details on the design rationale, implementation, and methods for data analysis. 

\subsection{General Overview}

\begin{figure*}[h!]
  \centering

 \includegraphics[width=0.9\linewidth]{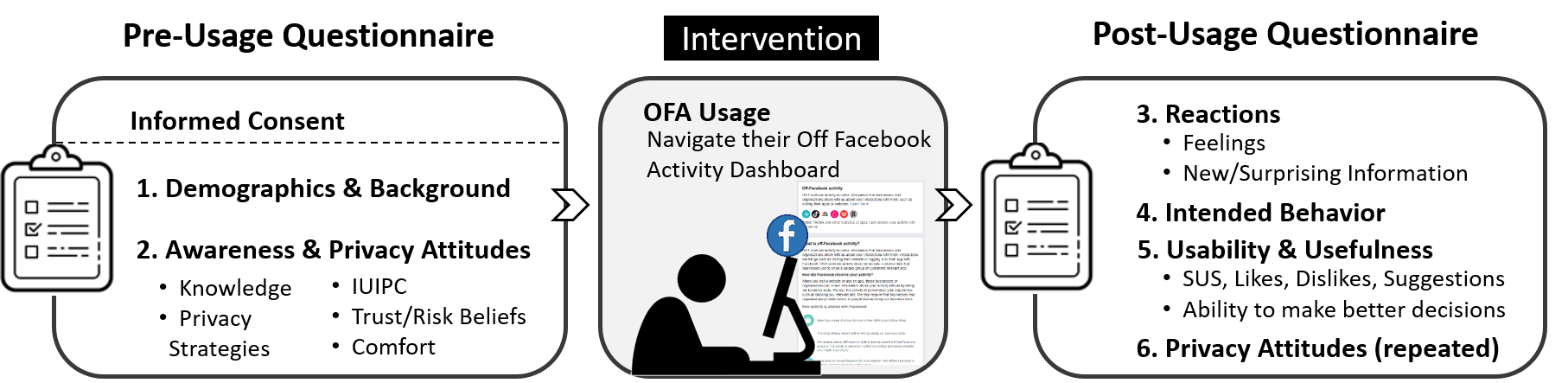}
 \caption{Structure of the user study. The study is divided into three parts: it starts with a questionnaire about initial awareness and attitudes, then users navigate their Off Facebook Activity (OFA) dashboard, and lastly they are questioned about their usage experience.}
    \label{fig:experimentOverview}
%  \Description{to add}
\end{figure*}

\textbf{Structure.} We designed an intervention study to be executed in three steps, summarized in Fig.~\ref{fig:experimentOverview}. First, participants fill a \textit{Pre-Usage Questionnaire} that registers their initial knowledge, privacy attitudes, and other demographic and background information. Second, they are instructed to use the ``Off-Facebook Activity'' transparency tool, from now on abbreviated as OFA. Third, participants answer a \textit{Post-Usage Questionnaire} to gauge the impact of the exposure to Facebook data collection practices after using OFA and  {evaluate the usability of the tool. Besides the survey, we complement and contextualize the participants' usability evaluation with an expert review of the tool conducted by the authors using a heuristics-based approach~\cite{lazar2017research} following Nielsen's 10 usability rules and usable privacy best practices~\cite{nielsen1990heuristic, lederer2004personal} }.

\textbf{Recruitment and Ethical Aspects.} We recruited participants through Mechanical Turk (MTurk)\footnote{\url{https://www.mturk.com/}}.  {We decided to focus on English-speaking respondents\footnote{ {English is the common language of the authors and used in the survey.}} in countries where GDPR was in place\footnote{ {To guarantee that users' data rights  regarding transparency are governed by this law, whose principles we use to discuss functionality and usability of the ``Off-Facebook Activity'' tool}} and with strong presence in MTurk \cite{MTurkDemographics}, hence we selected UK-based participants.} The study advertisement did not mention or even allude to privacy; it invited participants to a ``\textit{Facebook usage study}'' to ``\textit{understand how people use social networks these days}''. Participants were informed that the survey is anonymous and that all collected data is processed according to the EU General Data Protection Regulation (GDPR)~\cite{GDPR}, before asking for consent and confirmation of being over 18 years old. The questionnaires were administered via the LimeSurvey web-based survey tool\footnote{\url{https://www.limesurvey.org/}}, whose servers are located in Germany and comply with the European privacy regulations.
 The survey was approved by our university's Institutional Review Board.

\subsection{Survey Design}

Our survey includes a mix of quantitative questions, to measure privacy attitudes, behavior and usability, and open-ended qualitative questions, to get further insights on participants privacy mentalities. In the following we detail the most important question categories to address our research questions and refer the reader to the Appendix~\ref{app:survey} for the full questionnaire.

\begin{itemize}[leftmargin=*]
    \item \textbf{Demographics}. We seek to understand users backgrounds, more specifically: age, gender, education, technical knowledge, and frequency of FB usage. \textbf{\hyperref[q1]{Q1}-\hyperref[q6]{Q6}}. %\textcolor{teal}{nationality}

\item  \textbf{Awareness \& Privacy Attitudes}. To evaluate the level of initial awareness, we ask participants about their knowledge on FB data collection and what strategies they use to protect their privacy. Questions: \textbf{\hyperref[q26]{Q26}-\hyperref[q44]{Q44}}.

We evaluate user attitudes through the Internet Users' Information Privacy Concerns
(IUIPC) scale by Malhotra et al.~\cite{malhotra2004internet}. This construct allows to measure multidimensional facets of privacy, including attitudes towards the \textit{collection} of personal information, \textit{control} over personal information, and \textit{awareness} of privacy practices by online companies. We also borrow questions regarding trust and risk beliefs from Malhotra \& Agarwal~\cite{malhotra2004internet}, as these factors usually mediate privacy-decision making processes. Additionally, we ask participants about their level of comfort with data collection. To measure the effect of using the OFA transparency tool on privacy attitudes, we repeat this set of questions in the \textit{Pre-Usage} and \textit{Post-Usage} Questionnaires. Questions: \textbf{\hyperref[q7]{Q7}-\hyperref[q25]{Q25}}, and \textbf{\hyperref[q72]{Q72}-\hyperref[q72]{Q86}}. 

\item \textbf{Reactions \& Intended Behavior.} We use qualitative questions to complement the quantitative information on privacy attitude change, in order to capture additional details about how users react to FB data collection practices. Inspired by Weinshel et al.~\cite{weinshel2019oh}, we ask participants to report what new information and surprising information they discovered when using the OFA transparency tool. We also ask about their feelings post-usage.
 To explore intended behavioral changes, we ask participants to report how likely they are to take specific protective measures after using the dashboard. Questions: \textbf{\hyperref[q45]{Q45}-\hyperref[q54]{Q54}}.
		
\item \textbf{Usability \& Usefulness} To investigate the role of usability in transparency, we ask participants to quantitatively evaluate the OFA dashboard through the standard System Usability Scale (SUS)~\cite{brooke1996sus}, which has been widely used\footnote{Since SUS was developed by Brooke \cite{brooke1996sus} in 1996, more than 2300 individual surveys were conducted using SUS in over 200 studies by 2008 \cite{bangor2008empirical}} and proved valid and reliable to measure usability~\cite{peres2013validation}~\cite{kortum2014relationship}. To find out what features were most relevant or missing, we include open ended questions on likes, dislikes and suggestions for improvement. Additionally, we ask participants if the tool was useful for them to make better privacy decisions and how it shifted their risk-benefit perception on using Facebook.  Questions: \textbf{\hyperref[q55]{Q55}-\hyperref[q71]{Q71}}.

\end{itemize}

%%%%%%%%%%%%%%%%%%%%%%%%%%%%%%%%%%%%%%%%%%%%

\subsection{Analysis}

\textbf{Methods and Metrics.} Closed-ended responses were analyzed using targeted hypothesis testing with $\alpha$ =.05
, selecting the appropriate test based on the data type. We performed
 Wilcoxon signed-rank tests for repeated measurements (pre and post-OFA usage) on the Likert responses to IUIPC (\hyperref[q7]{Q7}-\hyperref[q16]{Q16}), trust and risk beliefs (\hyperref[q21]{Q21}-\hyperref[q21]{Q25}), and comfort (\hyperref[q18]{Q18})). In all cases, we first averaged responses across sub-scale items, treating the data as continuous. 

 Open-ended responses were analyzed following an iterative, inductive coding approach~\cite{inductiveCoding}. One member of the research team read responses and created the codebook with thematic codes, and a second researcher independently coded the full set of data. The inter-coder reliability for the final codes measured using Cohen's Kappa~\cite{cohen1960coefficient} was satisfactory for all questions ($\kappa$ > 0.7). 
  The cases where the coders differed in their final codes were discussed and reconciled.

\textbf{Pilot Testing.} Before publishing the study, we ran a pre-test with 9 subjects, asking them to provide feedback, and a pilot with 28 users. Based on these tests, we verified that the collected data was correct and in the expected format. The experience also helped us in understanding the expected completion time  {($\sim$30 min)}, and the feedback was useful to make minor modifications to the questionnaires regarding phrasing and order. Given the length of the study, we decided to include two attention check questions to filter bad quality responses from users that get fatigued and want to get away with the survey clicking trough it carelessly (\textbf{\hyperref[q24]{Q24}, \hyperref[q71]{Q71}}). Furthermore, to validate that the users actually opened the OFA dashboard, we asked them to upload an screenshot of the main interface, optionally showing the aggregated number of companies in their list, but including no personal information.  {Participants were given the possibility to donate their anonymous OFA data for enabling further research on data collection practices, and 6 of them did submit their information.}  The final version of the survey, which includes 87 questions, can be found in Appendix~\ref{app:survey}.
 
 {\textbf{Sample Size}. We determined the sample size with an \textit{a priori} power analysis using \texttt{G*Power}~\cite{faul2009statistical}. Accordingly, we observed the need for a sample size of N'= 57 for repeated measurement analysis with the Wilcoxon signed-rank tests to reach 95\% power  in detecting medium size effects (0.5), with alpha=0.05, two tailed. Hence, we set 57 users as a minimum requirement for our study for the quantitative analysis. Regarding the adequacy of this minimum for the qualitative analysis, research best practices recommend sample sizes of 10-20 participants~\cite{guest2006many} and usually below 50~\cite{below50}  to achieve saturation, i.e., the point at which no new themes emerge. As there can be variations depending on the domain of study, we targeted a larger sample, N = 100. We finally reached saturation after 35-40 responses. Our sample size also satisfies the recommendations for usability studies to discover most of the usability problems when evaluating an interface, which suggest 5-12 participants~\cite{hwang2010number, virzi1992refining} as a minimum. }

%% file: Results.tex
\label{sec:results}

 %\textbf{Participation.}
 The study was conducted in April 2021. Users were recruited through MTurk with the conditions of having a HIT approval rate $>$95\% and being located in the UK.  {We used MTurk premium filters to recruit verified FB account holders and have additional guarantee that the participants could access the tool.}
 
 We received 126 complete responses out of 172 submitted. The final sample, after filtering duplicates and answers that failed the check questions is N=100. It took participants, on average, around 32 minutes to complete the survey, and we paid them 10\$.  {This compensation scheme, slightly above average\footnote{ Minimum national wage in UK April 2021 was 8.91\pounds/hour \cite{minimumwageUK}, i.e., $\sim$11,13\$/hour, and we paid 20\$/hour }, was intended to factor in the additional effort of using a web tool and complete a long survey. A more detailed discussion on participants effort and mechanisms to ensure data quality is provided in the Limitations (Section \ref{sec:limitations}).}

 %\subsection{Demographics}

\textbf{Demographics.} The user sample is composed of a 38\%  of women, a 61\% of men, and 1 participant that preferred not to disclose their gender. Regarding age and education, participants are mostly young adults (72\% <35 years old) with a completed bachelor, professional, master, or doctorate degree (60\%). Other relevant factors are that we have a sample of active FB users (60\% use FB daily), and that 48\% of the participants have a background on IT. Table~\ref{tab:demographicsR} provides an overview of the age and gender distribution in our sample as compared to the UK Facebook population\footnote{\url{https://napoleoncat.com/stats/facebook-users-in-united_kingdom/2021/08/}}.
 A more granular overview of the sample characteristics can be found in Appendix~\ref{app:demographics}.
 %, \textcolor{red}{whose distribution is very similar to the general FB demographics \cite{demo1}. Add - Chisquared test for distribution of age and gender}
%in UK~\ref{FBdemographics}\textcolor{red}{addREf} .
%48% non IT bg 60 % every day 27 once a week

%%%%%%%%%%%%%%%%%%%%%%%%%%%%%
\input{table_demographics2}
%%%%%%%%%%%%%%%%%%%%%%%%%%%%%%%%

\subsection{RQ1: Awareness \& Privacy Attitudes}
\label{subsec:attitudes}

\textbf{Initial awareness.}
.
  Participants reported using FB mainly for socialization reasons (81\%), like staying in touch and sharing with friends and family. When it comes to privacy protection, the majority of users (99\%) do not have people they never met before as contacts, and 72\% of the participants configure their privacy settings. They generally adjust settings only once (40\%) and do not revisit the configuration frequently, but 11\% of respondents report changing privacy settings under specific events like \textit{``changes in the privacy policy''} or after a change in personal life circumstances, such as \textit{``after a breakup''}. 
  When we asked them about how they configure their \textit{privacy settings} and what other \textit{strategies} they use to protect their privacy (\hyperref[q37]{Q37}, \hyperref[q39]{Q39}), the general goal revealed in the answers is restricting access to their information by other persons.

  Consistently, 55\% of the participants reported using restricted lists to limit access to their FB profiles. 
  Other commonly reported strategies  
  include limiting the amount of information they post or have in their profiles (e.g., no real name or personal details) and the actions that others can take about their data, such as tagging and posting to their timeline. 
 
  Just a minority of the mentioned protections (9\% of the answers) were oriented at limiting tracking or its effects, including using tracker blockers, ad blockers, cookie managers, and not linking their FB account with other services. 
   Regarding \textit{knowledge} about the reasons why FB collects user data, even if the vast majority of users (92\%) reported being aware that FB gathers users' data for advertising purposes or marketing (\hyperref[q33]{Q33}), only 24\% of the participants had configured privacy settings to avoid being targeted by ads based on their activity within and/or outside the social network (\hyperref[q41]{Q41}). It is specially notable how accurately participants described how they think FB uses their data: 
   
   \begin{displayquote}
   \textit{``To create a profile of me that allows them to target me for advertising that I am likely to engage in. And also to create categories to fit each person into that can then be used for any number of future specific targeting.''} (P12)
   \end{displayquote}
   
   From participants' answers about their privacy awareness and behavior in FB, they seem to be well equipped to deal with social privacy but pay less attention to institutional privacy issues.

   While the reasons for inaction are unclear, it is plausible to think that restriction mechanisms to veto peer access are easier to understand and apply, as they resemble real life situations, but institutional tracking happens in the background, does not mirror real life situations, and therefore is not as straightforward to comprehend. Actually, from the 77 users who uploaded an OFA screenshot with a visible number of companies sharing their information with FB, 40\% underestimated the extent of tracking (\hyperref[q31]{Q31}).

   Before the study, the vast majority (85\%) never heard about OFA, and just 13 people had used it.    
   Almost all the participants (97) stated that they ``\textit{would like to use a system that shows you what information has been collected about you online.}'' In the following we analyse the impact of using such a tool, the OFA dashboard, on their attitudes.

\begin{figure}[t!]
  \centering

 \includegraphics[width=\linewidth]{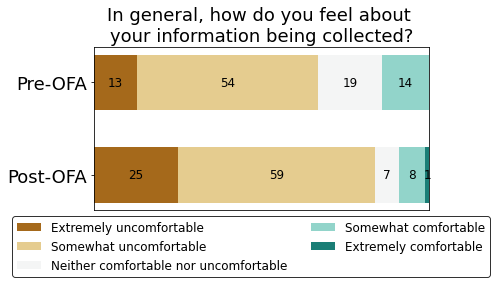}
 \caption{Proportions of the participants’ level of comfort about data collection before and after using the ``\textit{Off-Facebook Activity}'' (OFA) transparency tool.}
    \label{img:comfort}
\end{figure}

\textbf{Impact on attitudes.}
We measured the \textit{privacy attitudes} of participants through the IUIPC scale (\hyperref[q7]{Q7}-\hyperref[q16]{Q16}) before and after using the OFA tool. We also measured the \textit{trust and risk beliefs} (\hyperref[q21]{Q21}-\hyperref[q21]{Q25}), as well as the \textit{comfort} with data collection (\hyperref[q18]{Q18}) on Likert scales pre and post-exposure. To determine if there are significant changes, we performed Wilcoxon signed-rank tests. Results are summarized in Table~\ref{tab:trustrisk} and Fig. \ref{img:comfort}.  
 
%%%%%%%%%%%%%%%%%%%%
 \input{tableRQ1_}
 %%%%%%%%%%%%%%%%
%%%%%%%%%%%%%%%%%%%%%%%%%%%%%%%%%%%%%%%%%%
%comfort image
%%%%%%%%%%%%%%%%%%%%%%%%%%%%%%%%%%%%%%
%%%%%%%%%%%%%%%%%%%%%%%%%%%%%%%%%%%%%%%%%%%%%%%%%%%%%%%%%%%%%%%%

 We find that there is no effect on IUIPC. Subjects reported high levels of privacy concerns even prior to being exposed to transparency information. 

 When it comes to \textit{trust beliefs}, on the one hand we see a increase on trust post-usage regarding the perceived honesty of the social network ($Z = -2.89$, $p = 0.004$), with an effect size of $r = 0.35$, suggesting that this increase is moderate in size. On the other hand, trust regarding the predictable usage of users' information decreases $Z = 4.75$, $p < 0.001$), with a large effect size  ($r = $ 0.56). 
 
 With regard to \textit{risk beliefs}, we observe  no significant changes in risk perceptions. Users consider it risky to give their information to social networks, and this remains the same after interacting with the OFA tool.
 Additionally, participants' comfort with data collection decreased significantly ($Z = 3.023, p = 0.0024$) after exposure to transparency information with a moderate effect size ($r= 0.45$).

While the results on privacy concerns and risk do not show significant increase post-usage, this could relate to the fact that the scores were already very high initially. 
 The qualitative open reactions shed light on user changes in knowledge and sentiment and give further insights on why users' trust beliefs vary and why they are more uncomfortable after discovering the transparency information.

\subsection{RQ2: Reactions \& Intended Behavior}
\label{subsec:reactions}
\textbf{New and Surprising Information.} After using the assigned transparency tool, users reported what \textit{new knowledge} they gained (\hyperref[q45]{Q45}), as well as what information they found \textit{surprising} (\hyperref[q49]{Q49)}. Our qualitative analysis is summarized in Tables~\ref{tab:qualitative} and~\ref{tab:surprising}. 
\input{table_newInfo}
In most cases, the \textbf{new information} gained is related to the fact that there is \textit{tracking outside FB} (49.5\% of mentions) and to its extent (30.4\% of mentions). More specifically, users learn about the data sources, i.e., what are the apps and webs sharing their data, but not so much about the methods used for tracking and its purpose (only 7 participants report gaining this knowledge in both cases). Regarding the \textit{extent of tracking}, participants report learning about the amount of involved third parties and frequency of data collection. However, a lower amount of participants (n=7) report learning about the \textit{type of data} that is being collected. In fact, this is one of the main aspects where OFA should be improved according to user suggestions as we will detail in Section \ref{subsec:usability}.

Additionally, a 14.9\% of the coded responses reveal users learning new insights about \textit{privacy controls}. Some participants find out that there are settings to delete Off-Facebook activity and configure related ad preferences,  or they realize they can download a copy of their data. A minority of the users (5.15\% of the coded answers) learned about their own habits or nothing new. In this latter group, the majority were already OFA users.

\input{table_Surprising}

The percentage of users that found \textbf{surprising information} after using the OFA tool is 67\%.  
  Table~\ref{tab:surprising} summarizes the qualitative content analysis of participants' responses. Accordingly, surprising information falls into three categories.

  The most surprising information is by far the \textit{extent of tracking} (55.1\% of the codes), especially the amount of information collected and the number of third parties involved. This reaction suggests that users might not be aware of how consent to share these data is managed -- and several participants (n=5) alluded to this issue, for example: 
  
  \begin{displayquote}
 \textit{ ``I genuinely had no idea that so many websites and apps fed information back (sold my information?) to FB. For instance, one of my banks that I use an app for has a marketing app that sends info to FB. I mean really... FFS. If I can trust anyone, it should be my bank'' }	{(P175)}
  \end{displayquote}
  
    \begin{displayquote}
 \textit{ ``I didn't realise that if I visit a site, they can pass info to facebook. I assume it is because I have used the same browser to use both sites. I thought the list would be low, as I only use apps that ask ``link with facebook'' or similar, and I have said No to like 99\% of these. But it looks like websites don't do this. Well, they may have those privacy pop ups, but does any one read what they allow? I don't!'' (P85)}
     \end{displayquote}
 
  \begin{displayquote}
 \textit{If I don't explicitly realise or remember that I'm giving my permission to a company to share my business with them, I  should be able expect confidentiality in the same way I would from a public service such as a Library or Doctors Surgery. (P12)} 
 \end{displayquote}
    
  With regard to \textit{tracking outside FB} (36.73\% of the mentions), users are shocked to find unexpected or unknown applications and websites sharing their data. Some users are also surprised that FB has access to this information and that even offline activity and smartphone apps are tracked.

A minority of responses showed surprise about the availability of privacy controls (7.14\% of mentions) and one user was surprised in relation to discovering their own habits.

%%%%%%%%%%%%%%%%%%%%%%%%%%%%%%%%%%%%%%%%%%%%%%%%%%%%%%%
% Combined figure: feelings and intended behavior
%##################################################
\begin{figure*} [t!]
    
    \centering
    \subfigure[Feelings Post-OFA]{\includegraphics[width=0.55\textwidth]{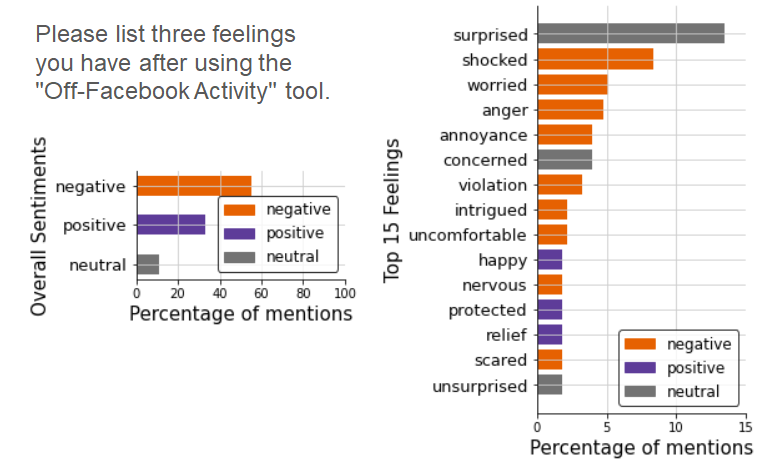}} 
    \label{fig:feelingsOFA}
    \hspace{-0.4cm}
    \subfigure[Intended Behavior Post-OFA]  {\includegraphics[width=0.45\textwidth]{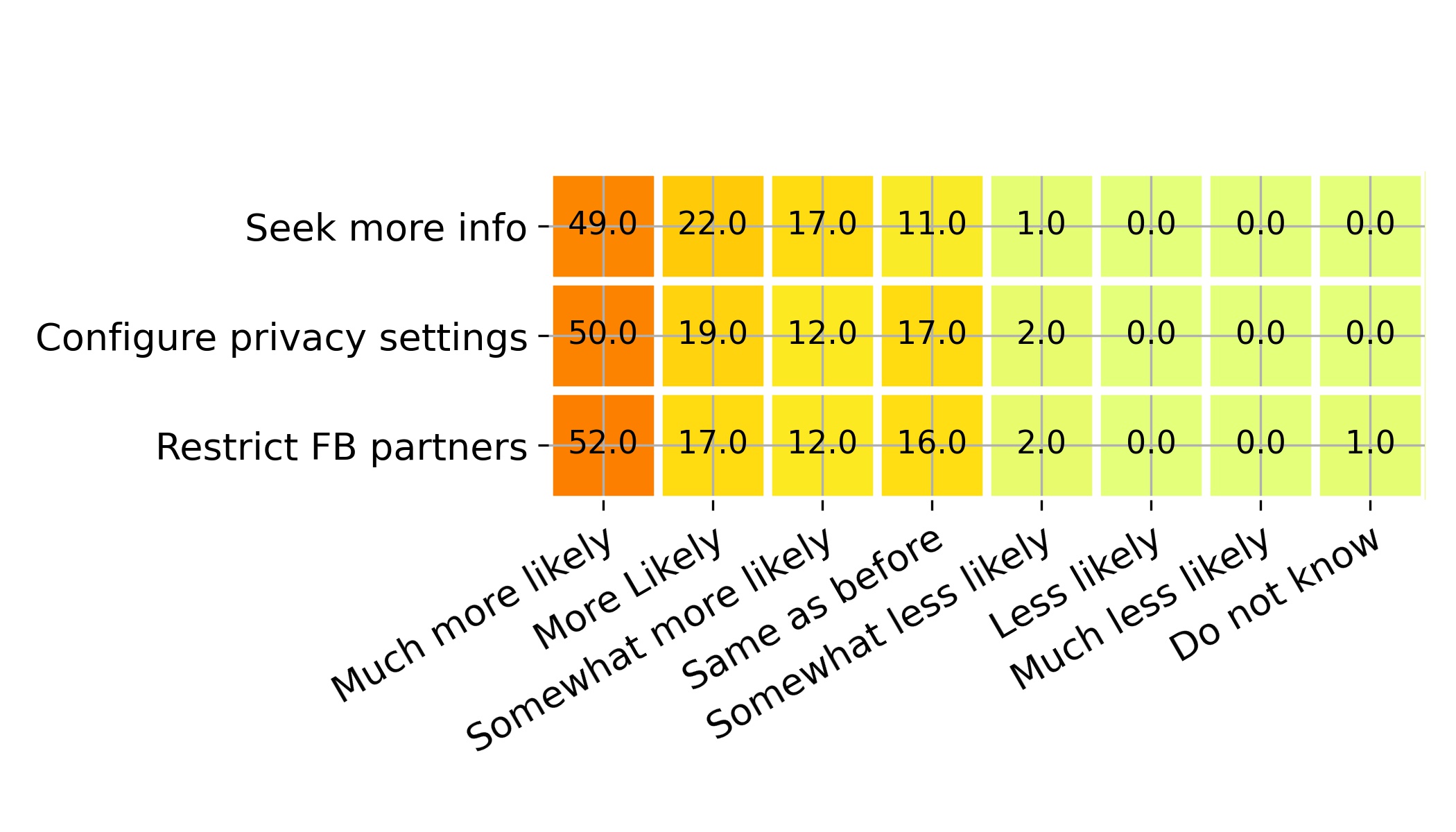}}\label{fig:behavior}

  \caption{ (a) Sentiment analysis (using NRC EmoLex~\cite{emolex}) of reported feelings after using the ``\textit{Off-Facebook Activity}'' (OFA) transparency tool. (b) Participants’ reported likelihood to take specific privacy protecting actions after using the OFA transparency tool. The heatmap shows the percentage of answers to a likelihood category (x-axis) for three privacy behaviors (y-axis). }
  \label{fig:feelingsBehavior}
     
\end{figure*}
%###############################################

\textbf{Feelings}. We conducted a sentiment analysis of users' feelings after using the OFA transparency tool. We specifically asked participants to list three feelings they had (\hyperref[q46]{Q46}).  We curated this list to remove invalid answers (e.g., long sentences instead of a word), corrected misspellings, and converted verbs and nouns to adjectives
 where necessary. The final list contained 275 words and 87 unique feelings. We processed this list using the NRC EmoLex lexicon~\cite{emolex} to cluster the feelings as positive, negative, or neutral.  
  The sentiment that dominates across all responses is negative (57,82\%), while the frequency of positive feelings in the whole set of responses is only 20\%. Fig. \ref{fig:feelingsBehavior} shows the top 15 feelings, which cover  $\sim$60\% of all user answers. As we can observe, 
   after being exposed to transparency information, the most common reaction was \textit{``surprise''} (13.45\%), a neutral feeling. Surprise is mostly triggered, as participants explained in a follow-up \textit{``Why do you feel like this?''} question (\hyperref[q47]{Q47}), because they underestimated the extent of FB tracking practices. This is supported by the quantitative observation in Section \ref{subsec:attitudes} about the difference in the number of companies listed by OFA and the user's preliminary estimation.   
  
    The dominant negative feelings in the top 5 are \textit{``shocked''}, \textit{``worried''}, and \textit{``anger''}.  These feelings come from the realization of the sheer amount of companies tracking them and the concerns about being invaded and violated, as well as preoccupation about how their data are used and shared.
  
  In turn, the small share of positive feelings (20\%) are related mainly to the gain of \textit{knowledge} and \textit{awareness} after using the transparency tool, \textit{curiosity} to explore further, or \textit{relief} to find out that they have options to control their privacy.
 
   It is to note that the majority (62\%) of users that reported positive feelings already used the OFA dashboard before, which suggests that the transparency and control that comes with a continued use of the dashboard are valuable. Participant P173, for example, felt horrified, shocked, and angry, but they clarify:
   
      \begin{displayquote}
 \textit{ `I don't currently feel this way but did the first time I accessed it. I think it is done in such a way that most normal people probably have no awareness about it. That is close to criminal in my estimation.''{(P173)}}
     \end{displayquote}

The prevalent negative feelings highlight the importance of raising awareness and designing transparency tools that are easy to navigate and that give clear actionable choices for users to control their data. We further elaborate on this topic during the discussion in Section~\ref{sec:discussion}. Now, we explore how exposure to OFA influences users' intentions to protect themselves.

\textbf{ {Intended} Behavioral Change}. In the post-usage questionnaires, we asked participants to rate how using the transparency tools had changed their likelihood to take three protective actions: \textit{i)} seek out more information about privacy settings in FB (\hyperref[q51]{Q51}); \textit{ii)} configure more private settings in FB (\hyperref[q52]{Q52}); and \textit{iii)} restrict FB partners sharing their information (\hyperref[q53]{Q53}). Figure~\ref{fig:feelingsBehavior} presents the results.  

Participants overwhelmingly reported increased intention to take all three privacy protective actions. Specifically, the percentage of OFA users that reported being between \textit{``somewhat more likely''} and \textit{``much more likely''} to take actions is: 88\% for seeking information, 81\% for configuring privacy, and 81\%  for restricting FB partners.

We also explored additional intended behavior with a follow-up question asking participants what measures would they take other than the three protective actions we specified. The majority of users reported they will explore and apply the settings offered by OFA to clear their history and manage their activity. Another common response was that they will modify their online behavior, for example, paying more attention to privacy policies, sharing less information with other apps, avoiding visiting certain webs or not using FB login. Some of the participants stated they would directly delete FB. Another significant observation is that several users reported they would search for similar transparency tools in other online social networks.

 Overall, these results signal people interests on transparency and, similar to Weinshel et al.'s research on web tracking~\cite{weinshel2019oh}, we observe the same trend that  comprehensive transparency information increases the reported privacy-preserving intentions,
  which underscores the value of TETs.

%%%%%%%%%%%%%%%%%%%%%%%%%%%%%%%%%%%%%%%%%%
 % IMAGE USEFULNESS: BETTER DECISIONS
 %%%%%%%%%%%%%%%%%%%%%%%%%%%%%%%%%%%%%%
 \begin{figure}[t!]
  \centering
  \includegraphics[width=\linewidth]{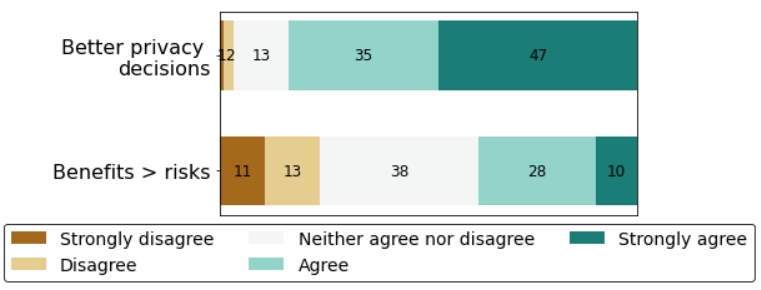}
  \caption{Participants' agreement levels to: \textit{``feeling able to make better privacy decisions''} and considering that \textit{``benefits of using Facebook outweigh the risks''}, after using the ``\textit{Off-Facebook Activity}'' (OFA) transparency tool.}
  \label{fig:usefulness}
\end{figure} 
%%%%%%%%%%%%%%%%%%%%%%%%%%%%%%%%%%%%%%%%%%%%%%%%%%

%#######################################
% SUBSECTION USABILITY AND USEFULNESS
%%%%%%%%%%%%%%%%%%%%%%%%%%%%%%%%%%%%%%%%%

\subsection{RQ3: Usability \& Usefulness}
\label{subsec:usability}

\textbf{Usability.} According to ISO 9241~\cite{iso19989241111998en}, usability is the extent to which a software tool ``\textit{can be used by specified users to achieve specified goals with effectiveness, efficiency, and satisfaction in a specified context of use}''. We understand the goals for a transparency tool as in the GDPR description that individuals should know ``\textit{[what] personal data concerning them are collected, used, consulted or otherwise processed and to what extent the personal data are or will be processed}'' \cite{GDPR}.

 Considering these definitions, we measure and analyze the satisfaction usability dimension of the OFA tool through the SUS scale ({\hyperref[q58]{Q58}-\hyperref[q67]{Q67}}) and use open-ended questions on likes, dislikes, and suggested improvements ({\hyperref[q68]{Q68}-\hyperref[q70]{Q70}}) to capture richer information on how the tool supports the transparency goal.
 
 Participants rated the OFA tool with a\textit{ SUS score} of $69.65$ ($\pm 16.47$). This value is close to, but not yet at the ``acceptable'' usability threshold of 70 points based on previous research~\cite{bangor2008empirical},  { which indicates that there is a need for improvement.}
 
  {Previous to analyzing the answers to participants' open questions, we conducted an expert heuristic review~\cite{nielsen1990heuristic, lazar2017research} of the OFA user interface, observing three main areas where the tool is limited:}
 
 \begin{figure}[t!]
  \centering

 \includegraphics[width=\linewidth]{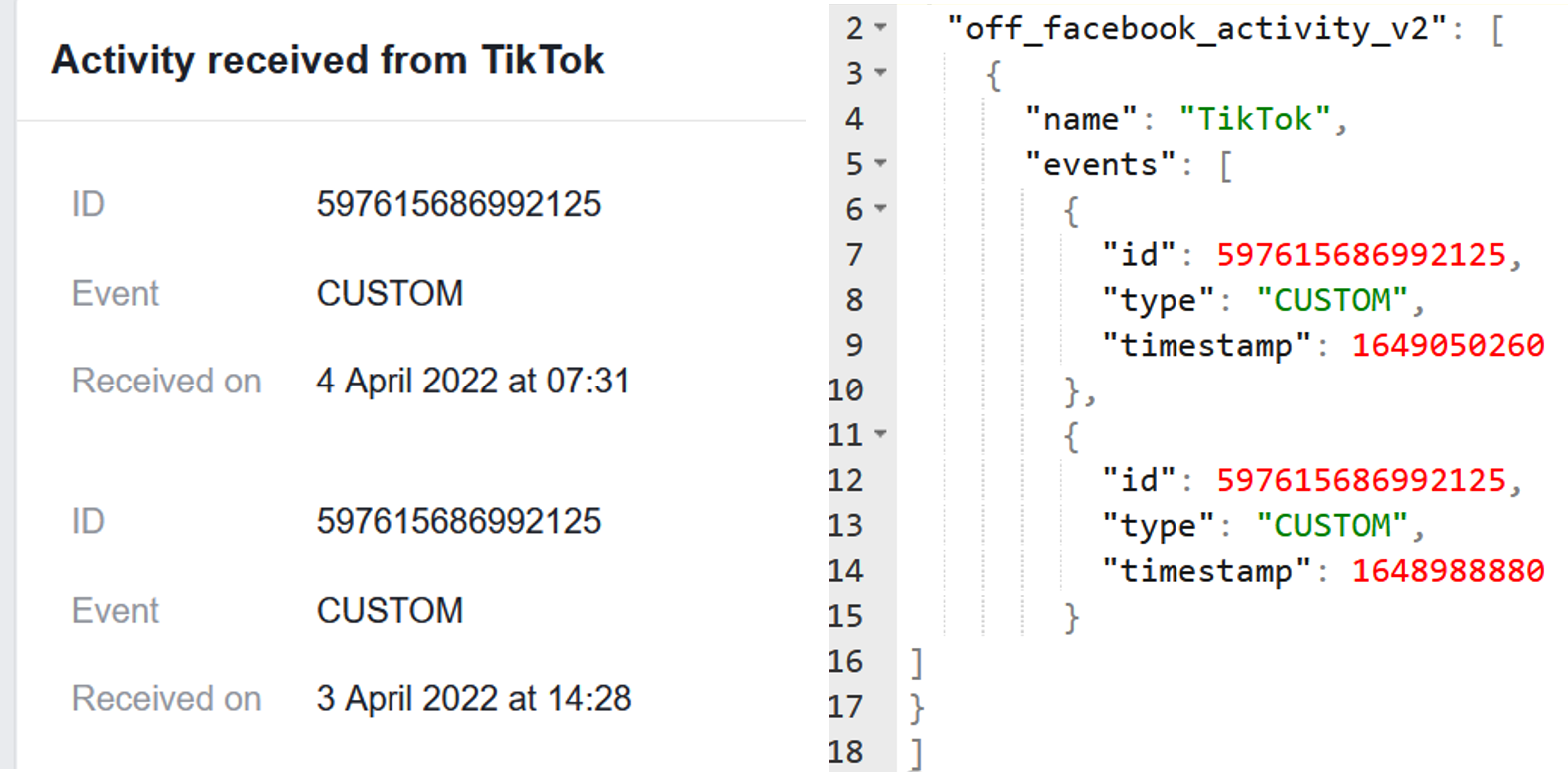}
 \caption{Example excerpt of detailed activity found in an HTML (left) and JSON (file) downloaded using the ``\textit{Off-Facebook Activity}'' tool.}
    \label{fig:OFARaw}
\end{figure}
 
 \begin{enumerate}
     \item  {\textbf{Transparency}. The level of transparency detail is vague. The OFA interface displays a list of third parties that sent data to Facebook (see Fig.~\ref{fig:interfaces}). However, when clicking on a specific entry to get further details it only shows generic information, the same for every entry, providing examples of possible interactions, such as searches or purchases. But these examples constitute a limited subset of the data types captured by FB. For example, the set of events that can be shared through the Facebook Pixel\footnote{https://developers.facebook.com/docs/facebook-pixel/reference} specifies 10 additional categories, including information of when a person books an appointment to visit a concrete location (e.g., a shop). Besides, companies can use the more opaque ``CUSTOM'' event to send anything they consider relevant that it is not captured in a standard event\footnote{CUSTOM events constituted 20\% of the events and one of the three most common types shared for the 6 participants that donated their anonymized data in our pilot study
}. If the user
wants to know the specific data exchanged, they are referred to a download
page, where they can obtain a copy of their Off-Facebook activity, but it is not straightforward to discover and navigate the Off-Facebook data export. Furthermore, the download takes time  and users may lose motivation due to the lack of immediacy. Even for users that have the time and ability to go through the data export, they will find scattered raw data as lists of timestamped event types per application (see Fig.\ref{fig:OFARaw}), from which is difficult to extract meaningful conclusions and, in many cases, CUSTOM events with no indication about the exact exchanged data. }

\item   {\textbf{Control}. OFA provides limited functionality for users to control their data, considering the intervenavility rights covered in the GDPR. It allows users to detach their Off-Facebook activity information from their Facebook account, but not to delete it (right to erasure). Users cannot withdraw consent to stop the data sharing flow (right to withdraw consent). On the positive side, OFA does allow users to get a copy of their data (right to access), and to provide feedback on their data, though not to rectify it (right to rectification).}

\item   {\textbf{Presentation}. The presentation of both transparency information and control options to users falls short in providing usable and clear visualizations, instructions, and feedback. Users must scroll and wait for the web page to be refreshed in order
to explore the full list of companies that shared their information, which is tedious and not practical, specially when the list is long. There is no option to quickly visualize all companies at once and see how they compare to each other in terms of tracking. Though sorting options are available, the results are always shown in a \textit{scroll to refresh} list. Furthermore, the naming of control options to detach past and future activity do not convey the fact that the information is not erased; there is a 
 clear button accompanied by a standard trash can icon, usually representing the act of deletion. These different terminology and inconsistent iconography go against usability best practices as they can lead to confusion and mislead users. Similarly, instructions to download and navigate the Off-Facebook activity are not explicit. While the tool allows to give feedback about the collected activity (unrecognized, misuse, inappropriate, other), there is no information about how this labeling is used by Facebook or how it is useful for the user, and the labeling status is not shown anywhere in the interface. }
  \end{enumerate}
  
  {We used the above three categories of usability issues as main themes to group and contextualize users' responses about the tool usability in terms of likes, dislikes, and suggestions. The results of the qualitative analysis are summarized in Table~\ref{tab:usabilityAnalysis}.}

 \input{tableUsability.tex}

 With regard to users' \textbf{likes}, participants positively evaluated that FB \textit{provides transparency} %( {16.9\% of mentions})
 and shows information about \textit{companies and interactions} involved in user data sharing ( {27.9\% of mentions combined}). They also liked that there are options to \textit{clear their history}, \textit{download} their data, and \textit{manage future activity} ( {21.8\% of mentions combined}). Furthermore, some users praised the simplicity of the \textit{interface} and how \textit{easy to use and understand} it is ( {31,3\% mentions combined}). On the negative side, the \textbf{dislikes} highlighted important limitations in the transparency and control functionalities. The main transparency problems are that the \textit{level of detail} of the information is not enough ( {10.1\% of mentions}), for example regarding how it was collected, in which user device, and what was the exact exchanged data. With regard to privacy controls, participants criticized the excessive effort required to remove \textit{consent} ( {5.3\% of mentions}) 
  %configure the settings instead of having a single on/off toggle to deactivate all data collection
  . Several users mentioned that the process should be opt-in rather than opt-out, and other participants noticed that the offered controls just detach the collected data from the user account but do not actually stop the collection and processing, which is undesirable. This is an interesting point to further investigate as it is not clear if all users realize that the OFA controls do not end the information flow and what are the implications. Finally, the most frequently highlighted concern ( {18.1\% of mentions}) is the lack of \textit{awareness} about the tool existence and how difficult it is to locate, which might impact users' trust perceptions:
 
    \begin{displayquote}
  {\textit{`` It should be more obvious to find this tool as I have never heard of it before today'' (P64)}}
    \end{displayquote}

     \begin{displayquote}
 \textit{``I find it dishonest that FB separates its privacy settings into different places, with too many screens to click through.'' (P45)}
    \end{displayquote}
 
   {These aspects related to consent and awareness were not discovered during the heuristic review previous to the user study, which highlights the value of combining both methodologies.} Additionally, while many users did not dislike specific aspects of the OFA tool, they remarked their discomfort with their data being shared:
 
     \begin{displayquote}
 \textit{ ``I didn't dislike anything about the tool, but I disliked that all this information was being shared in the first place.'' (P114)}
     \end{displayquote}
 
\begin{displayquote}
 \textit{ ``I disliked realising how much my privacy has been compromised!'' (P160)}
     \end{displayquote}

   The most relevant participants' \textbf{suggestions} to improve the tool are aligned with the disliked aspects (see Table~\ref{tab:usabilityAnalysis}) and we will summarize the main action points that can be derived from them in our discussion in Section ~\ref{sec:discussion}.

\textbf{Usefulness.} To further investigate if users feel actually more capable of making better privacy decisions after using the OFA dashboard, we directly asked them. The results, together with participants' evaluation if the benefits of FB outweigh the risks, are shown in Fig.~\ref{fig:usefulness}.
An 83.3\% of the participants report feeling capable of making better privacy decisions after using OFA. 

As for the risk-benefits tradeoff, the results show that only a minority of users agree that risks outweigh the benefits. However we do not have additional information on how they actually made this evaluation. 
 In this regard, it would be interesting to analyze in future studies to what extent transparency helps in understanding and judging risks. 

Despite the majority of users feel more capable of making privacy decisions post-usage, free answers about intended behavior still convey the complexity involved in this type of decisions (it is not only Facebook, but an ecosystem of applications involved in tracking), the lack of clear privacy controls, and the feeling of resignation that privacy is a price to pay:

\begin{displayquote}
\textit{``I won’t take any other action because I don’t know how to, I feel hopeless''}{(P67)}
\end{displayquote}

\begin{displayquote}
\textit{``Over 500 listed, I am in shock, how do I remove all of these without having to go through one by one?!''}
{(P76)}
\end{displayquote}

\begin{displayquote}
\textit{``I have turned off the feature, and cleared the history, but I am aware that FB will still use such data for ad targeting if it can collect it.''} {(p45)}     
\end{displayquote}

In the light of these result, we argue that, to further empower users making privacy decisions, transparency tools should be better designed to communicate risk and provide easy, actionable, and meaningful privacy-protective options.

%% file: table_demographics2.tex
% \usepackage{booktabs}
% \usepackage{multirow}
\begin{table}[]
\caption{Participant demographics}
\small
 %\footnotesize
\centering
\begin{tabular}{@{}llcc@{}}
\toprule
                  &&  \textbf{Study Sample}  &  \textbf{UK FB Users} \\
                  &&(\textit{n = 100})& (\textit{n = 49.69Million})\\
                  \midrule
                  
\multirow{3}{*}{\rotatebox[origin=c]{90}{\textbf{Gender}}} & Female& 38 \%& 52.9 \%\\ 
                   & Male & 61\%& 47.1\% \\
                   &No Answer  &1\% & - \\
\midrule
\multirow{7}{*}{\rotatebox[origin=c]{90}{\textbf{Age}}}           
                    &13-17 & -& 4.2\%\\
                    & 18-24 &29\%& 17.1\%  \\
                 &  25-34& 43\%& 24.6\%  \\
                  &  35-44&16\%& 18.7\% \\
                  &  45-54& 8\%& 14.9 \\
                  &  55-64& 2\% &11.1\%\\
                  &  >= 65 &  1\%& 9.5\%\\ 
                  & No Answer &1\%& -\\
\bottomrule
%\cmidrule(l){2-3} 
\end{tabular}
  \label{tab:demographicsR}
\end{table}

%% file: tableRQ1_.tex
%!!!Version without Comfort

% \usepackage{booktabs}
% \usepackage{multirow}
%patri: add footnote to mention that full questions are listed in the appendix
\begin{table}[t!]
\caption{Wilcoxon signed rank tests for IUIPC-based privacy concerns, trust and risk beliefs before and after using the ``Off Facebook Activity'' (OFA) transparency tool. Responses were given on a 7-point Likert scale from ``strongly disagree'' (1) to ``strongly Agree'' (7).}

 %\small
 \footnotesize 	
 \centering

 \begin{tabular}{@{}clcccccc@{}}

\toprule
%\multicolumn{6}{c}{ \textbf{Impact on Privacy Attitudes}}             \\ \midrule
                   &  & \multicolumn{2}{l}{\textbf{Pre-OFA}} & \multicolumn{2}{l}{\textbf{Post-OFA}} && \\
                  &  & Avg.  & SD & Avg. & SD & \textbf{Z}& \textbf{\textit{p}}\\
                 % \cline{3-4} \cline{5-6}
                 \midrule
\multirow{4}{*}{\rotatebox[origin=c]{90}{\textbf{IUIPC}}} & Awareness  & 6.29  &  0.69 & 6.35    & 0.77 & -0.99 & 0.318 \\
                  & Control  & 5.70 & 0.93& 5.81   & 0.79  & -0.69& 0.238\\
                  &  Collection &5.92 & 0.74 &  6.01  & 0.79 &-1.184& 0.236\\
                  \cline{3-6}
                  &  Total & 5.97  &0.55& 6.06   &0.61  &-1.88&0.059\\
                  \midrule
\multirow{2}{*}{\rotatebox[origin=c]{90}{\textbf{Trust}}} & Honesty & 3.31 & 1.59 & 3.37 &1.77 & -2.89 & %\cellcolor{green!25}0.004  \\
 \cellcolor{green}0.004  \\
                  & Predictability  &4.29  &1.56  & 3.81 &1.82 & 4.75& %\cellcolor{green!25}$<$.001 \\
                  \cellcolor{green}$<$.001 \\
\midrule

\multirow{2}{*}{\rotatebox[origin=c]{90}{\textbf{Risk}}} 
                  & Privacy Loss & 5.57 & 1.04 & 5.72 & 1.02  & -1.02& 0.309\\
                  & General Risk & 5.08 &1.33  & 5.37 &1.09&-1.89& 0.058 \\
                  \bottomrule
\end{tabular}
\label{tab:trustrisk}
\end{table}

%% file: table_NewInfo.tex
%%%%%%%%%%%%%%%%%%%%%%%%%%%%%%%%%%%%%%%%%%%%%%%%%%

% \usepackage{multirow}
% \usepackage{booktabs}

\begin{table*}
\caption{Categories and codes used to code free text answers on new information gained after using the ``\textit{Off-Facebook Activity}'' transparency tool. The frequency for each category appears within parentheses. The last two columns indicate the number of times and percentage a code appears in the dataset. %The inter-coder reliability measured with the Cohen's Kappa is $\kappa$ =0.94.
}% when the category was coded.} %Multiple codes could be applied per category.}
\label{tab:qualitative}

\footnotesize
\centering
\begin{tabular}{p{2cm}p{2.4cm}p{10.5cm}p{0.2cm}p{0.5cm}} 
\toprule
\textbf{Category}                       & \textbf{Codes} & \textbf{Representative Quote} &  \textbf{N} & \textbf{\%}  \\\midrule
\multirow{2}{*}{\adjustbox{stack=ll}{\textbf{Tracking Off Facebook} (49.5\%)}}     &   Tracking Sources              & ``I found out what apps and websites are sharing my activities'' 
                      & 83 & 42.8\% \\ 
&  \cellcolor{Gray} Tracking Methods  & \cellcolor{Gray} ``How your information is shared when you buy something: they use business tools.'' 
                    & \cellcolor{Gray}7 &\cellcolor{Gray} 3.6\% \\

&   Tracking Purpose  &  ``That the information is used to learn about your interests and offer you personalised adverts.''&   6& 3.1\% \\
            \midrule
\multirow{4}{*}{\adjustbox{stack=ll}{\textbf{Extent of Tracking }(30.4\%)}} &   \cellcolor{Gray}Amount of Data &\cellcolor{Gray} ``I learned there is a vast amount of data being collected and transmitted about me.'' &\cellcolor{Gray} 15     &\cellcolor{Gray} 7.7\%\\

&\parbox[t]{2.4cm} {Amount of\\ Webs/Apps}   &  ``I realised that over 1,000 applications use and have access to my data.'' & 36 & 18.6\%\\ %%%

&   \cellcolor{Gray} Data Persistence &\cellcolor{Gray} ``My data is stored for ever seemingly, which is disturbing.'' &\cellcolor{Gray}    1 &\cellcolor{Gray}0.5\%\\

&   Type of Tracked Data & ``I now know that basically every app I use will send information of each time I have logged in and out of the app, or added an item to a wishlist or basket to Facebook.'' &7     &3.6\%\\

\midrule

\parbox[t]{2cm}{\textbf{Privacy \\Control} (14.9\%)} &\cellcolor{Gray} Controls Available & \cellcolor{Gray}`I learned that it's possible to clear this history and presumably no longer get relevant ads from those companies, until I visit their site again.  I also learned that it's possible to manage which off-Facebook activity is saved.'' &\cellcolor{Gray}     26&\cellcolor{Gray} 13.4\%\\
&Data Available & ``That I can download my recorded activity'' &     3&1.5\%\\

\midrule
\multirow{2}{*}{\adjustbox{stack=ll}{\textbf{Other} (5.15\%)}} 
&\cellcolor{Gray} Nothing & \cellcolor{Gray}``Didn't gain any new information.'' &\cellcolor{Gray}    6&\cellcolor{Gray}3.1\%\\
& Own Habits  & ``Knowing more what I use.'' &   4  &2.1\%\\

\bottomrule
\end{tabular}

\end{table*}

%% file: table_Surprising.tex
%%%%%%%%%%%%%%%%%%%%%%%%%%%%%%%%%%%%%%%%%%%%%%%%%%

% \usepackage{multirow}
% \usepackage{booktabs}

\begin{table*}
\caption{Categories and codes used to code free text answers on surprising information gained after using the ``\textit{Off-Facebook Activity}'' transparency tool. The frequency for each category appears within parentheses. The last two columns indicate the number of times and percentage a code appears in the dataset. %The inter-coder reliability measured with the Cohen's Kappa is $\kappa$ =0.8.
}% when the category was coded.} %Multiple codes could be applied per category.}
\label{tab:surprising}

\footnotesize
\centering
\begin{tabular}{p{2cm}p{2.4cm}p{10.5cm}p{0.2cm}p{0.5cm}} 
\toprule
\textbf{Category}                       & \textbf{Codes} & \textbf{Representative Quote} &  {\textbf{N}} & {\textbf{\%}}  \\\midrule

\multirow{5}{*}{\adjustbox{stack=ll}{\textbf{Tracking Off Facebook} (36.73\%)}} 
& Unexpected Webs/Apps & ``They have access to apps that are supposed to be completely secure, like online banking ones.'' & 
9& 9.18\%\\

&\cellcolor{Gray}  Unknown Webs/Apps    & \cellcolor{Gray} ``The most concerning element is that there are sites I don't even recognise which is worrying.'' &  \cellcolor{Gray}6  &6.12\%  \cellcolor{Gray} \\

&FB has Access to Data     &  ``I did not know that Facebook collected all this information!'' &   12& 12.24\%\\

&\cellcolor{Gray}  Apps are tracked & \cellcolor{Gray} ``This is surprising because I didn't expect all these apps installed on my phone to be sharing this type of information with Facebook.'' & \cellcolor{Gray}  3&\cellcolor{Gray}3.06\% \\

& Tracking Methods  & ``It is a bit surprising that it shows sites that I have used in google without signing in or login.'' & 
6& 6.12\%\\

\midrule
%%%%%%%%%%%%%%%%%%%%%%%%%%%%%%%%%%%%%%%%%%%%%%%%%%%%%%%55
\multirow{4}{*}{\adjustbox{stack=ll}{\textbf{Extent of Tracking} (55.1\%)}} &\cellcolor{Gray}Amount of Data   & \cellcolor{Gray}``I feel surprised that so much information is being sent to one place.'' &\cellcolor{Gray}17     &\cellcolor{Gray}17.35\%\\

& \parbox[t]{2.4cm}{Amount of\\ Webs/Apps}       &  ``I found it surprising how many websites had transmitted my information and stored it.'' &   27& 27.55\% \\ %%%

& \cellcolor{Gray}Data Persistence   & \cellcolor{Gray}``How long the data is kept for.'' &\cellcolor{Gray}2     &\cellcolor{Gray}2.04\%\\

& Type of Tracked Data &  ``I was surprised that they have all this information including date and time of visit'' & 3   & 3.06\%  
\\
&\cellcolor{Gray}\parbox[t]{2.4cm}{ Unconsented\\ Exchange} & \cellcolor{Gray}``These other apps are sharing my information with Facebook when I haven't expressly consented to sharing this information.'' &\cellcolor{Gray}  5 & \cellcolor{Gray} 5.1\%\\

\midrule
\multirow{2}{*}{\adjustbox{stack=ll}{\parbox[t]{2cm}{\textbf{Privacy \\Control}} (7.14\%)}} 
& Data Available & ``Open availability of the information.'' &     3&3.06\%\\
&\cellcolor{Gray} Controls Available & \cellcolor{Gray}``I learnt how I can manage my off Facebook activities and how to disconnect it.'' &\cellcolor{Gray}1     &\cellcolor{Gray}1.02\%\\
& Public Unawareness & ``How this is not at all public (or publicly known)'' &     3&3.06\%\\

\midrule
\multirow{1}{*}{\adjustbox{stack=ll}{\textbf{Other} (1.22\%)}} 
& \cellcolor{Gray}Own Habits &\cellcolor{Gray} ``I didn't know that I was sharing my activity with that many websites.'' &     \cellcolor{Gray}1&\cellcolor{Gray}1.02\%\\
\bottomrule
\end{tabular}

\end{table*}

%% file: tableUsability.tex
%%%%%%%%%%%%%%%%%%%%%%%%%%%%%%%%%%%%%%%%%%%%%%%%%%
% \usepackage{multirow}
% \usepackage{booktabs}

\begin{table*}
\caption{Categories and codes used to code free text answers on likes, dislikes, and suggestions given by participants after using the ``\textit{Off-Facebook Activity}'' transparency tool. The columns N and \% indicate the number of times and percentage a code appears in the respective dataset. Related codes in the likes/dislikes/suggestions are shown in the same row to facilitate interpretation.} %The inter-coder reliability for each question, measured with the Cohen's Kappa, is $\kappa$ =0.89, 0.88 and 0.94.
% when the category was coded.} %Multiple codes could be applied per category.}
\label{tab:usabilityAnalysis}

\footnotesize
\centering
\begin{tabular}{p{1.5cm}p{3.4cm}p{0.2cm}p{0.5cm}p{3.4cm}p{0.2cm}p{0.5cm}p{3.4cm}p{0.2cm}p{0.5cm}} 
\toprule
%(100 Responses, 148 Codes), (90 Responses, 93 Codes), (93 Responses, 96 Codes)
& \multicolumn{3}{l}{\textbf{Likes }} & \multicolumn{3}{l}{\textbf{Dislikes  }} & \multicolumn{3}{l}{\textbf{Suggestions }}\\

\textbf{Category}  & \textbf{Code} & \textbf{N} &  \textbf{\%} & \textbf{Code} & \textbf{N} &  \textbf{\%}& \textbf{Code} & \textbf{N} &  \textbf{\%}  \\\midrule
\multirow{6}{*}{\adjustbox{stack=ll}{\textbf{Transparency}}}   & shows companies \& interactions &22&15\%&&&&&& \\ 
&  \cellcolor{Gray}  level of detail &\cellcolor{Gray} 1&\cellcolor{Gray}0.7\%&\cellcolor{Gray}level of detail&\cellcolor{Gray}9&\cellcolor{Gray}9.6\%&\cellcolor{Gray}increase detail&\cellcolor{Gray}11&\cellcolor{Gray}11\% \\ 

& purpose explained &1&0.7\%&&&&explain purpose&1&1\% \\ 

&  \cellcolor{Gray}  tracking method explained 
&\cellcolor{Gray}1&0.7\%\cellcolor{Gray}&\cellcolor{Gray}&\cellcolor{Gray}&\cellcolor{Gray}&\cellcolor{Gray}explain tracking method&\cellcolor{Gray}1&\cellcolor{Gray}1\% \\ 

& security information &1&0.7\%&&&&&& \\ 

&  \cellcolor{Gray}  transparency provided &\cellcolor{Gray}19&\cellcolor{Gray}12.9\%&\cellcolor{Gray}&\cellcolor{Gray}&\cellcolor{Gray}&\cellcolor{Gray}&\cellcolor{Gray}&\cellcolor{Gray} \\ 

\midrule

\multirow{6}{*}{\adjustbox{stack=ll}{\textbf{Control}}}  
& download data &5&3.4\%&&&&&& \\ 
&\cellcolor{Gray} clear history &\cellcolor{Gray}10&\cellcolor{Gray}6.8\%&\cellcolor{Gray} no coarse grained clear history\cellcolor{Gray}&\cellcolor{Gray}2&\cellcolor{Gray}2.1\%&\cellcolor{Gray}add coarse-grained clear&\cellcolor{Gray}1 &1\%\cellcolor{Gray}\\
& manage future activity &17&11.6\%&no coarse-grained manage future&1&1.1\%&&& \\ 
&\cellcolor{Gray} controls available &\cellcolor{Gray}8&\cellcolor{Gray}5.4\%&\cellcolor{Gray}&\cellcolor{Gray}&\cellcolor{Gray}&\cellcolor{Gray}&\cellcolor{Gray} &\cellcolor{Gray}\\
& &&&(un-)consent mechanism &5&5.3\%&easy (un-)consent &    12&12\% \\ 
&\cellcolor{Gray} &\cellcolor{Gray} &\cellcolor{Gray}&no real delete\cellcolor{Gray}&1\cellcolor{Gray}&1.1\%\cellcolor{Gray}&\cellcolor{Gray}&\cellcolor{Gray}&\cellcolor{Gray} \\

\midrule

\multirow{3}{*}{\adjustbox{stack=ll}{\textbf{Presentation}}}  
& easy to use/understand &32&21.8\%&difficult to use/understand&7&7.4    \%&make it clear/easier&3&3\% \\ 
&\cellcolor{Gray} efficient &4\cellcolor{Gray} &\cellcolor{Gray}2.7\%&\cellcolor{Gray}&\cellcolor{Gray}&\cellcolor{Gray}&\cellcolor{Gray}&\cellcolor{Gray}&\cellcolor{Gray} \\
& interface &14&9.5\%&interface&7&7.4\%&improve interface visualizations&10&10\% \\ 

\midrule

\multirow{3}{*}{\adjustbox{stack=ll}{\textbf{Other}}}  
&\cellcolor{Gray}nothing &\cellcolor{Gray}10&\cellcolor{Gray}6.8\%&\cellcolor{Gray}nothing&\cellcolor{Gray}45&\cellcolor{Gray}47.9\%&\cellcolor{Gray}nothing&\cellcolor{Gray}41&\cellcolor{Gray}41\% \\ 
&good for advertisers  &1&0.7\%&&&&& \\ 
&\cellcolor{Gray}  password protected &\cellcolor{Gray}1&\cellcolor{Gray}0.7\%&\cellcolor{Gray}&\cellcolor{Gray}&\cellcolor{Gray}&\cellcolor{Gray}&\cellcolor{Gray}&\cellcolor{Gray} \\
& && &awareness&17&18.1\%&awareness&19&19\% \\ 
&\cellcolor{Gray}   &\cellcolor{Gray} &\cellcolor{Gray}&\cellcolor{Gray}&\cellcolor{Gray}&\cellcolor{Gray}&\cellcolor{Gray}more regulation&\cellcolor{Gray}1&\cellcolor{Gray}1\% \\

\bottomrule
\end{tabular}

\end{table*}

%% file: Discussion.tex
\label{sec:discussion}
%POSTMAJOR: \textcolor{red}{ easily identifiable through a recognizable security icon} 
Our study contributes to a deeper understanding on the importance of transparency about online mass surveillance. We showed that when users are enabled to learn how their data are collected, they are eager to do so and they discover new insights that shift their intention to take protective actions.  Even if we measured intended behavior, recent events, like the massive migration of Whatsapp users to Signal after the former  published a privacy policy update implying unnecessary data sharing with FB~\cite{signal}, highlight the value of transparency to change behavior. Transparency tools also bring value to well-informed, privacy-active users, as these tools can help verifying that data collection and processing is done according to their expectations, i.e., their controls are working. Our findings  strengthen previous work on web tracking and ad explanations transparency, where users were also moved to apply protective measures~\cite{weinshel2019oh} after navigating their data.

 Contextualizing our results within the state of the art, we identify two critical issues to be addressed in the OFA transparency tool, and potentially applicable for other transparency dashboards: 1) current levels of transparency are not enough, and 2) users' real choices to control their data are limited.  {Additionally, usability of  transparency interfaces should be improved to better support users. In the following, we discuss concrete problems and suggestions for future research and TETs implementations based on our user study and related work.}

\subsection{Minimal transparency is not enough}

The principle of transparency as stated in the GDPR requires that: 

\begin{displayquote}
\textit{``Any information addressed to the public or to the data subject be concise, easily accessible and easy to understand, and that clear and plain language and, additionally, where appropriate, visualisation be used''.} 
\end{displayquote}

The OFA tool, as other TETs in the literature, provides limited, ambiguous, opaque, and difficult-to-access information. %The only data that is easy to access is a list of applications and web sites that shared unspecified user information, but more granular detail requires users to download files and interpret the information themselves, which research shows is.
 {Getting more granular details requires users to download raw data exports, which are difficult to navigate and interpret, and need improvement to be useful~\cite{veys2021pursuing}. }

Accuracy is a problem too, as the FB interface warns users that it receives \textit{``more details and activity than what appears here''}.  {Participants showed their discomfort regarding the fact that not all data are shown and pointed out that the warnings about it should be more visible.} 
For effective transparency, more accurate and comprehensive information is required and users want these details. This finding echoes previous studies in the related area of ad transparency showing that users prefer detailed  information~\cite{wei2020twitter, eslami2018communicating}.  {Furthermore, besides concrete information about data exchange, the OFA tool falls short at providing other pieces of information required by the GDPR, namely: information about risks and consequences,  clear purpose of data collection and usage, and how the data is processed (e.g., inferences). }
Additionally, transparency should be usable. The current usability score for the OFA tool is marginally acceptable according to the standard SUS scale.  In particular, the importance of visualizations in usable transparency communication is also highlighted in the GDPR \cite{GDPR} and previous work on tracking transparency confirmed that users obtain the most value from graphics \cite{weinshel2019oh}, making it an interesting element to include and to explore in further research.
%, users feel generally more informed and are more likely to protect themselves.
 It would be interesting to study usability patterns for effective transparency communication, considering other transparency domains and well-researched guidelines and privacy iconography, such as the ``privacy nutrition labels''~\cite{emami2020ask}. Additionally, participants in our user study suggested to improve OFA with notifications. It remains to be studied if users would benefit from this type of real time or ``in-context'' approach to transparency, i.e., informing when the sharing occurs and not only as an ex-post all-in-one-place list of trackers.  {Furthermore, one of the core findings in this study is the lack of awareness and the desire of people to get knowledgeable.} It is vital to raise awareness of transparency tools, transparency rights, and media competence of the general public. More than a year after the OFA tool was realised, the vast majority of our participants have not heard about it. Indeed, as they suggested, the tool could be made easier to locate and frequently advertised.

\subsection{Do users really have control over their data?}
%Exercising data control is hard}
%Actionable controls are missing . Opting out is complicated 
%No real choice
The GDPR also reads that for transparency:

\begin{displayquote}
\textit{``Natural persons should be made aware of risks, rules, safeguards and rights in relation to the processing of personal data and how to exercise their rights in relation to such processing''.}
\end{displayquote}
The current state of transparency in FB, also illustrated in related work in different domains, sheds doubts over how easy is for users to actually exercise their rights regarding data control. While OFA provides means to take privacy actions, they are quite limited, specially regarding data erasure and consent: users can only disconnect the data from their account\footnote{This is not even possible for all 3rd parties, e.g., Facebook does not allow to detach data coming from Oculus VR headsets and services.} %and restrict ads based on personal data,
 but they can not delete the data or stop data collection from the interface.  Furthermore, when detaching their personal data, users are only warned about the loss of functionality that the decision might entail (e.g., same amount but less relevant ads) but privacy risks are not communicated. Our participants were surprised to see unexpected applications and webs sharing their data, which suggests that current consent notices are not working well, as otherwise users would be aware of their approval to share data. Additionally, opting out is overwhelmingly complex, as it has been pointed out also in previous research~\cite{habib2020s}. In our pilot study, the 6 participants that uploaded their anonymized data, had on average 425 companies sharing their information with the social network. Reading the privacy policies of each of them to understand how to restrict data sharing and sending formal requests to ask for it, would require an excessive and unmanageable amount of time~\cite{mcdonald2008cost, obar2020biggest}. Even if a highly motivated user takes the time and effort to do this, it is only a tiny part of the overall surveillance picture, so their actions can feel irrelevant 
%to protect privacy in the whole online tracking ecosystem
, which leaves users with no real path to choose privacy. Participants suggested on/off toggles to allow for efficient batch consent withdrawal. We need standardized protocols to enable and automate this functionality. Engagement from research, technologists, and policymakers is required to devise new ways to put users in control.

%%%%%%%%%%%%%%%%%%%%%%%%%%%
%Design recommendations
%%%%%%%%%%%%%%%%%%%%%%%%%%%%%%%

%%%%%%%%%%%%%%%%%%%%%%%%%%%%%%%%%%%%%%%%%%%%%%%%%%%%%%%%%%%%%%LIMITATIONS
%%%%%%%%%%%%%%%%%%%%%%%%%%%%%%%%%%%%%%%%%%%%%%%%%%%%%%%%%%%%%

\subsection{Limitations}
\label{sec:limitations}
Our study is to be interpreted considering its limitations. Like many survey-based studies, our results suffer from self-report biases. To limit desirability bias (respondents misreporting their answers to make themselves look better)~\cite{kreuter2008social}, we did not mention that the study focus was on privacy until the debriefing information provided at the end of the survey.  {We asked demographic questions at the beginning of the survey to increase response rates, but this might have introduced stereotype biases~\cite{stereotypebias}}. Since we used a survey rather than an interview, we did not have the opportunity to follow up with participants regarding answers we found interesting.  {However, our initial results can be a starting point for designing interview or observation-based studies to further explore how users interact with the native FB transparency tool or alternative designs.}

 {
To guarantee the quality of the collected data, we followed best practices for conducting research with MTurk~\cite{peer2014reputation},  using high-reputation workers, additional filters to target FB account holders, and requesting proof of usage of the transparency tool. Additionally, we implemented attention checks and filtered participant answers with implausible response times. While the duration of the study can be considered long, fairness in task time advertising and payment play an important role in increasing data quality even for long surveys~\cite{MTurkFairness}, so we opted for an above average payment as an incentive. We also piloted with a large number of users to get an accurate estimate of the time to complete the survey, so the advertised time was strongly aligned with the actual time taken by our participants. According to existing research~\cite{survey20min, survey20min2}, survey respondents get fatigued after 20 minutes, which might impact the data. In our case, however, there were two questionnaires with a break in between, in which participants had to use the tool under evaluation. Our participants spent an average of 11 minutes in the pre-usage survey, 9 minutes using the tool, and 12 minutes answering the post-usage survey, which yields a total average time of 23 minutes answering questions in a non-continuous manner, potentially having less impact on fatigue. In fact, checking the average time effort per question in the pre and post surveys, we did not observe a reduced time investment in the post-usage questions, but it was consistent across the whole study and aligned to expected average times per question~\cite{ikart2019survey}. }

Regarding the sample,  { we targeted participants located in UK, recruited through Mechanical Turk}. It is known that MTurk workers are generally younger and more privacy-sensitive than the the overall population  {\cite{xia2017our, kang2014privacy}}, which can be a reason why the privacy concerns are very high from the beginning. Studies with other populations and a more representative sample are thus desirable.

 Furthermore, to investigate behavioral reactions to transparency,  we focus on self-reported qualitative feedback about intended future behavior.   { We explore this dimension as a proxy to actual privacy behavior because intentions are a prerequisite~\cite{intendedbehavior} to take action. Though correlated, the results are to be understood as an upper bound to actual behavior, which would be affected by multiple other factors, such as usability \cite{acquisti2020secrets}. }Follow-up studies on actual behavior modification would be required to understand if action follows intention and to better understand mediating factors hindering or facilitating change.

%% file: Conclusion.tex
\label{sec:conclusions}
%\textcolor{red}{TO-DO}
In contrast to common belief that citizens were increasingly careless about their privacy, users indeed have a vital interest in the ways their data are collected and processed. 
They even adapt when given the opportunity to research this information with suitable transparency tools.
%Contrary to the usual portray of users as privacy careless, people are willing to know how their data are collected and used. 
Our study on FB underlines the importance of institutional transparency for users to feel informed and capable of making privacy decisions and giving informed consent to the processing of their data. 
Indeed, it illustrates that current consent notices are inefficient: users were mostly surprised to discover to which extent third party companies are supplying their data to Facebook, despite the fact that they supposedly gave their fully informed consent to this practice. 
We conjecture that online transparency requires improvement, as the information asymmetry is going to put users at increased disadvantage and severely affects their right to self-determination. This is bound to exacerbate the current surveillance economy and to negatively impact free, and democratic societies. 
We posit that it is necessary to raise awareness about existing transparency tools towards this end. 
In addition, there is demand to develop new, usable tools that provide both detailed information and insight towards potential privacy impact, as well as effective means for users to exercise control over their data.

%% file: Survey.tex
\label{app:survey}

\subsection{Pre-Usage Questionnaire
}

\subsubsection{Demographics and Background}
\small
\begin{itemize}
\item [Q1]\label{q1} With what gender do you identify?
  \textcolor[rgb]{0.5,0.5,0.5}{(\Square Male \Square Female \Square Non-binary \Square Other \Square Prefer not to say)}
\item [Q2] How old are you?
\textcolor[rgb]{0.5,0.5,0.5}{( \Square 18-24\Square 25-34  \Square 35-44  \Square 45-54 \Square 55-64 \Square 65 or older  \Square Prefer not to say)}
\item [Q3] What is the highest degree or level of education you have completed? \textcolor[rgb]{0.5,0.5,0.5}{( \Square Some high school \Square High school \Square Some college \Square Trade, technical, or vocational training  \Square Bachelor’s degree  \Square Master’s degree  \Square Professional degree  \Square Some graduate School  \Square  Doctorate  \Square Prefer not to say)}
 \item [Q4] What nationality do you most identify with? 
 \item [Q5] Which of the following best describes your educational background or job field?
\textcolor[rgb]{0.5,0.5,0.5}{( \Square I have an education in, or work in, the field of computer science, engineering, or IT. \Square I do not have an education in, or work in, the field of computer science, engineering, or IT. \Square Prefer not to say.)}
\item [Q6]\label{q6} Approximately, how often do you access your Facebook account? \textcolor[rgb]{0.5,0.5,0.5}{( \Square Every day  \Square  A few times a week \Square  About once a week \Square A few times a month \Square Once a month  \Square Less than once a month)}

\end{itemize}

\subsubsection{Privacy Attitudes}
\small
\textbf{IUIPC Awareness, Collection and Control Subscales}\\
\textcolor[rgb]{0.5,0.5,0.5}{(Answer choices:  \Square Strongly agree  \Square Agree \Square Somewhat Agree  \Square Neither agree nor disagree  \Square Somewhat Disagree \Square Disagree  \Square Strongly disagree)}
\begin{itemize}
\item[Q7]\label{q7}  Companies seeking information online should disclose the way the data are collected, processed, and used. 
\item[Q8] A good consumer online privacy policy should have a clear and conspicuous disclosure.
\item[Q9] It is very important to me that I am aware and knowledgeable about how my
personal information will be used.
%\end{itemize}%\textbf{IUIPC Collection Subscale}\\
%\textcolor[rgb]{0.5,0.5,0.5}{(Answer choices:  \Square Strongly agree  \Square Agree \Square Somewhat Agree  \Square Neither agree nor disagree  \Square Somewhat Disagree \Square Disagree  \Square Strongly disagree)}\\
%\begin{itemize}
%
%\item [•]
\item[Q10] It usually bothers me when online companies ask me for personal information.
\item[Q11] When online companies ask me for personal information, I sometimes think
twice before providing it.
\item[Q12] It bothers me to give personal information to so many online companies.
\item[Q13] \label{q13}I’m concerned that online companies are collecting too much personal information about me.
%\end{itemize}
%
%\textbf{IUIPC Control Subscale}\\
%\textcolor[rgb]{0.5,0.5,0.5}{(Answer choices:  \Square Strongly agree  \Square Agree \Square Somewhat Agree  \Square Neither agree nor disagree  \Square Somewhat Disagree \Square Disagree  \Square Strongly disagree)}\\
%
%\begin{itemize}
    \item [Q14] Consumer online privacy is really a matter of users’ right to exercise control and autonomy over decisions about how their information is collected, used, and shared.
    \item [Q15] Consumer control of personal information lies at the heart of consumer privacy.
    \item [Q16] \label{q16} You believe that online privacy is invaded when control is lost or unwillingly reduced as a result of a marketing transaction.
\end{itemize}

\bigskip
\textbf{General Attitudes}
\begin{itemize}
\item [Q17] In your own words, how do you explain privacy?
\item [Q18] \label{q18} In general, how do you feel about your information being collected? \textcolor[rgb]{0.5,0.5,0.5}{(Answer choices:  \Square Extremely uncomfortable, \Square Somewhat uncomfortable,
\Square Neither comfortable nor uncomfortable, \Square Somewhat comfortable \Square Extremely comfortable)}
 \item [Q19] You would only provide accurate and personal information at SNSs if their control policy is verified / monitored by a reputable third party.\textcolor[rgb]{0.5,0.5,0.5}{(Answer choices:  \Square Strongly agree  \Square Agree  \Square Neither agree nor disagree  \Square Disagree  \Square Strongly disagree)}
 \item [Q20]\label{q20} If it were available, you would like to use a system that shows you what information has been collected about you online.\textcolor[rgb]{0.5,0.5,0.5}{(Answer choices:  \Square Strongly agree  \Square Agree  \Square Neither agree nor disagree  \Square Disagree  \Square Strongly disagree)}
%IUIPC extra awareness question  (could work as a check)
%Social network sites (SNSs) should disclose the way the data are collected, used and processed.
\end{itemize}

\bigskip
\textbf{Trust and Risk Beliefs}\\
\textcolor[rgb]{0.5,0.5,0.5}{(Answer choices:  \Square Strongly agree  \Square Agree  \Square Neither agree nor disagree  \Square Disagree  \Square Strongly disagree)}

\begin{itemize}
    \item [Q21] \label{q21} SNSs are in general predictable and consistent regarding the usage of the information.
    \item [Q22] SNSs are always honest with users when it comes to using the information that users would provide.
    \item [Q23] In general, it would be risky for a user when he/she gives his/her personal information in SNSs.
     \item [Q24]\label{q24} We use this question to discard the answers of people who are not reading the questions. Please select “Strongly Agree” to preserve your answers. \textcolor{olive}{[\textit{Check question}]}
    \item [Q25] \label{q25} There would be high potential for privacy loss associated with giving
personal information to SNSs.
\end{itemize}

\subsubsection{Awareness} 
\textbf{Experience with Off-Facebook Activity
}
\begin{itemize}
    \item [Q26]\label{q26} Have you ever heard about the  ``Off-Facebook Activity'' privacy tool?\textcolor[rgb]{0.5,0.5,0.5}{ (\Square Yes \Square No)}
    \item  [Q27] \textcolor[rgb]{0.5,0.5,0.5}{If Q26=Yes:} Please list the information that you already knew about the “Off-Facebook Activity” tool.
\item [Q28] Have you ever used the ``Off-Facebook Activity''  tool?\textcolor[rgb]{0.5,0.5,0.5}{(\Square Yes \Square No)}
\item  [Q29] \textcolor[rgb]{0.5,0.5,0.5}{If Q28=Yes:} How did you use the “Off-Facebook Activity” tool? \textcolor[rgb]{0.5,0.5,0.5}{(\Square You have cleared history of your activities \Square You have disconnected your activities from your Facebook account by turning of the ``Off-Facebook Activity'' privacy tool \Square You have downloaded your activities from your Facebook account  \Square You have accessed the information that Facebook has about you)}
\item  [Q30] \textcolor[rgb]{0.5,0.5,0.5}{If Q28=Yes:}   Do you like to clear your ``Off-Facebook Activity'' for privacy reasons?\textcolor[rgb]{0.5,0.5,0.5}{(\Square Yes \Square No)}

\end{itemize}

%\subsubsection{Privacy Awareness}
%\subsubsection{Awareness:
\bigskip
\textbf{Knowledge about Facebook Data Collection}
%including what information users think facebook collects, with which frequency and for which purposes, but also how do they feel about it
%about the info: oline purchases and opening smartphone app are not so evident, let's see of their opinion changes after the tool
\begin{itemize}
    \item [Q31] \label{q31} How many companies do you think that share their information about you with Facebook? \textcolor[rgb]{0.5,0.5,0.5}{(\Square 0 companies \Square Less than 20 companies \Square 20-60 \Square 60-100 \Square More than 100 companies)}

\item [Q32]  Which types of information do you think Facebook collects as you use it? \textcolor[rgb]{0.5,0.5,0.5}{(\Square Websites that you visit \Square Every advertisement topic you have clicked \Square A list of all companies that has your contact information from the advertisements you have clicked \Square All of your contact information from your phone book \Square Your online purchases \Square Your exact location \Square Every social event you are invited to through Facebook \Square Your personal information (name, email, gender, ...)
 \Square When you use an app in your smartphone \Square Other)}

\item [Q33] \label{q33} In your own words, please describe the purposes for which you think the information you selected above is collected.

\end{itemize}

%\subsubsection{Facebook Privacy Strategies}
%\subsubsection{Awareness: 
\bigskip
\textbf{Privacy Behavior in Facebook}
%explore how users protect their privacy in FAcebook, social and institutional privacy settings and other strategies, ad ecosystem awareness and behavior
\begin{itemize}
	\item [Q34] Who are your friends on Facebook? \textcolor[rgb]{0.5,0.5,0.5}{(\Square Friends \Square Friends of friends \Square Professors \Square Colleagues \Square Boss \Square Parents/uncles/aunts \Square People you have never met \Square Patients \Square Other)}
	\item [Q35] Do you use restricted lists to limit access to your profile?\textcolor[rgb]{0.5,0.5,0.5}{(\Square Yes \Square No)}
	\item [Q36] Do you use your Privacy Settings on Facebook?\textcolor[rgb]{0.5,0.5,0.5}{(\Square Yes \Square No)}
	\item [Q37]\label{q37}  \textcolor[rgb]{0.5,0.5,0.5}{If Q36=Yes:} How do you configure  Facebook settings to protect your privacy?
	\item [Q38] \textcolor[rgb]{0.5,0.5,0.5}{If Q36=Yes:} How often do you change your Privacy Settings on Facebook?\textcolor[rgb]{0.5,0.5,0.5}{(\Square Never \Square I did it once Frequently \Square When something changes (e.g., when you see a policy change notification, when you add a new friend, when you enter a relationship, etc). Please specify)}
	\item [Q39] \label{q39} 	If you follow any other strategy to protect your privacy in Facebook, please briefly explain it. (For example: not using your real name to avoid being searchable, configuring Tinder so that you are not discoverable for Facebook friends, not tagging your pictures)
	\item [Q40] Have you ever looked at your Facebook ad preferences?\textcolor[rgb]{0.5,0.5,0.5}{(\Square Yes \Square No )}
	\item [Q41] \label{q41} Have you ever configured Privacy Settings related to Facebook advertisements?\textcolor[rgb]{0.5,0.5,0.5}{(\Square Yes \Square No )}
	\item [Q42]  \textcolor[rgb]{0.5,0.5,0.5}{If Q41=Yes:} How did you change your ads Settings?\textcolor[rgb]{0.5,0.5,0.5}{(\Square 
	You do not allow ads based on the data from Facebook partners \Square You do not allow ads based on your activity on Facebook company product that you see elsewhere \Square You do not allow ads that include your social actions \Square Other)}
\item [Q43]  	Do you follow any strategies to prevent Facebook from collecting information about you?\textcolor[rgb]{0.5,0.5,0.5}{(\Square Yes \Square No)}
\item [Q44] \label{q44} \textcolor[rgb]{0.5,0.5,0.5}{If Q43=Yes:} Please explain these strategies.
		
\end{itemize}

%\subsection{Usage Instructions}
%Now, please access your Facebook account on a computer and spend some time exploring the Off-Facebook activity tool. You
%can access it as follows:
%Click on the top right icon of your Facebook profile . Select ``Settings & Privacy'' > ``Settings''. Click on ``Your Facebook
%Information'' on the left column. Click on ``Off-Facebook Activity''. From this screen, you can also click ``Manage Your Off-
%Facebook Activity'' for more information. You'll be asked to re-enter your password.
%If you could not find the option, here is a direct link: https://www.facebook.com/off_facebook_activity/

\subsection{Intervention}

\begin{figure}[h!]
  \centering
 \includegraphics[width=0.9\linewidth]{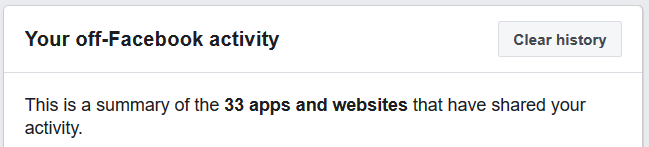}
 \caption{Sample Screenshot of the Off-Facebook Activity tool}
    \label{img:screenshot}
\end{figure}

Now, please access your Facebook account on a computer and access the Off-Facebook Activity tool following these steps:

\begin{itemize}
    \item Click on the top right icon of your Facebook profile. Select ``Settings $\&$ Privacy'' $>$ ``Settings''.
    \item Click on ``Your Facebook Information'' on the left column.
    \item Click on ``Off-Facebook Activity''. From this screen, you can also click ``Manage Your Off-Facebook Activity'' for more information. You'll be asked to re-enter your password.
\end{itemize}
If you could not find the option, here is a direct link: \url{https://www.facebook.com/off_facebook_activity/}

Once you access the Off Facebook Activity interface, please take a screenshot of the summary information. It should look similar to the image in Figure \ref{img:screenshot}.

%%%%%%%%%%%%%%%%%%%%%%%%%%%%%%%%%%
%%%%%%%%%%%%%%%%%%%%%%%%%%%%%%%%%

%   consists of a pre- and post-usage questionnaire with overall 100 questions.
%96 plus 4 checks
%The chart in figure 3.1 shows the subgroups of questions in both pre- and post-usage questionnaire. The questions in this survey are structured in three different ways: as openended questions, closed-ended questions with ordered responses categories, or closed-endedquestions with unordered response categories [57].

Don't include any personal information. Please \textbf{spend some time using and exploring the Off-Facebook Activity tool} and come back to answer the remaining questions about your experience.

\subsection{Post-Usage Questionnaire}
%After receiving instructions to access the Off-Facebook Activity tool, users are told to spend some time exploring the functionality and respond to the Post-Usage questionnaire as follows.

\subsubsection{Open-Ended Reactions}
%novelty, nobody explored feelings before(?)
\begin{itemize}
	\item  [Q45] \label{q45}  Please list the new information that you gained by using this tool. 
  \item  [Q46] \label{q46} Please list three feelings you have after using the "Off-Facebook Activity" tool.
  \item [Q47] \label{q47} Why do you feel this way? (Please explain for each of the listed feelings.)
  \item  [Q48] Did you find any surprising information by using this tool? \textcolor[rgb]{0.5,0.5,0.5}{(\Square Yes \Square No)}
  \item  [Q49]  \label{q49} \textcolor[rgb]{0.5,0.5,0.5}{If Q48=Yes:} Please explain these surprising information in your own words.
	\item  [Q50]   Do you have any question about what you saw in the tool?

\end{itemize}

\subsubsection{Post-Usage Intended Behavior}
\small

Compared to before you used the tool,

\begin{itemize}
	\item  [Q51] \label{q51}  How likely are you to seek out more information about possible privacy-related settings on Facebook?
\item  [Q52] \label{q52}  How likely are you to configure more private settings on your Facebook?
\item  [Q53] \label{q53} How likely are you to restrict Facebook’s partners sharing your information?
\textcolor[rgb]{0.5,0.5,0.5}{Answer choices for all questions: \Square Much more likely \Square More likely \Square Somewhat
more likely \Square About the same as before \Square Somewhat less likely \Square Less likely \Square Much less likely \Square  Don’t know
}\end{itemize}
\begin{itemize}
\item  [Q54] \label{q54} Please, mention any other action that you have taken or will take to protect your privacy as a result of using this tool.
\end{itemize}

\subsubsection{Usefulness}
\small
\begin{itemize}
\item  [Q55] \label{q55}Compared to before you used the tool, you feel that you are able to make better decisions about your privacy-related actions.\textcolor[rgb]{0.5,0.5,0.5}{(\Square Strongly agree  \Square Agree  \Square Neither agree nor disagree  \Square Disagree  \Square Strongly disagree)}

\item  [Q56] \label{q56} Why do you use Facebook?
\item  [Q57] After using the tool, you think the benefits of using Facebook outweigh the privacy risks. \textcolor[rgb]{0.5,0.5,0.5}{(\Square Strongly agree  \Square Agree  \Square Neither agree nor disagree  \Square Disagree  \Square Strongly disagree)}
\end{itemize}

\subsubsection{Usability}
\small
~\\
\textbf{System Usability Scale (SUS)}.
During the rest of this survey, we use the term ``system'' to refer the Off-Facebook activity tool. Next, we show statements
about your experiences with the system. Please select the answer choice that best describes your agreement.
%or disagreementwith these statements.
\textcolor[rgb]{0.5,0.5,0.5}{(Answer choices:  \Square Strongly agree  \Square Agree  \Square Neither agree nor disagree  \Square Disagree  \Square Strongly disagree)}
%The post-study starts with the SUS questionnaire, items SUS01-SUS10, answered with a 5-point Likert-Scale from \textit{Strongly Disagree} to \textit{Strongly Agree}. Additionally, we added questions PQ1 and PQ2 to understand users preferences and intention on continued use.
%and follows with these two additional questions:

\begin{itemize}
\item [Q58] \label{q58}I think that I would like to use this system frequently
\item [Q59] I found the system unnecessarily complex
\item [Q60] I thought the system was easy to use
\item [Q61] I think that I would need the support of a technical person to be able to use this system
\item [Q62] I found the various functions in this system were well integrated
\item [Q63] I thought there was too much inconsistency in this system
\item [Q64] I would imagine that most people would learn to use this system very quickly
\item [Q65] I found the system very cumbersome to use
\item [Q66] I felt very confident using the system
\item [Q67] \label{q67}I needed to learn a lot of things before I could get going with this system

\end{itemize}

\textbf{Likes, Dislikes and Suggestions}

\begin{itemize}
\item [Q68] \label{q68} List the aspects, if any, that you liked about the Off-Facebook Activity\textcolor[rgb]{0.5,0.5,0.5}{$/$Facebook Data Visualization} tool.
\item [Q69] List the aspects, if any, that you disliked about the Off-Facebook Activity\textcolor[rgb]{0.5,0.5,0.5}{$/$Facebook Data Visualization} tool.
\item [Q70] \label{q70} Do you have any suggestion(s) to improve the Off-Facebook Activity\textcolor[rgb]{0.5,0.5,0.5}{$/$Facebook Data Visualization} tool? If so, please elaborate below.
\item [Q71] \label{q71} We use this question to discard the answers of people who are not reading the questions. Please select “Strongly Agree” to preserve your answers. \textcolor{olive}{[\textit{Check question}]}
\end{itemize}

\subsubsection{Post-Usage Privacy Attitudes and Awareness}
%~\\
\begin{itemize}[leftmargin=1.3cm]
    \item [Q72-Q86] \label{q72} We repeat questions Q7-Q16, Q18, Q21-Q23, and Q25.% from the Pre-Usage survey.
%\subsection{Survey 2 Instrument}
%\label{app:survey2}
\end{itemize}

\subsubsection{General Feedback Question}
%~\\
\begin{itemize}

\item [Q87] \label{q87} Is there anything else you would like to add about Facebook privacy practices or this survey in general? 
\end{itemize}

%%%%%%%%%%%%%%%%%%%%%%%%%%%%%%%%%%%%%%%%%%%%%5

\section{Detailed Demographics}
\label{app:demographics}

% \usepackage{booktabs}
% \usepackage{multirow}
\begin{table}[hb!]
\caption{Detailed participant demographics}
%\small
 %\footnotesize
\centering
\begin{tabular}{@{}llc@{}}
\toprule
                  %&&  \textbf{Study Sample}  \\
                  &&\textit{n = 100}\\
                  \midrule
                  
\multirow{3}{*}{\rotatebox[origin=c]{90}{\textbf{Gender}}} & Female& 38 \%\\ 
                   & Male & 61\%\\
                   &No Answer  &1\% \\
\midrule
\multirow{7}{*}{\rotatebox[origin=c]{90}{\textbf{Age}}}           
                    & 18-24 &29\% \\
                 &  25-34& 43\%\\
                  &  35-44&16\% \\
                  &  45-54& 8\% \\
                  &  55-64& 2\%\\
                  &  >= 65 &  1\%\\ 
                  & Prefer not to say &1\% \\
\midrule
\multirow{9}{*}{\rotatebox[origin=c]{90}{\textbf{Education}}}                  
&Some high school      &                       1\\
&High school            &                     10\\
&Some college            &                    24\\
&Bachelor’s degree        &                   38\\
&Trade, technical, or vocational training&     4\\
&Professional degree                      &    2\\
&Master’s degree                           &  18\\
&Doctorate                                  &  2\\
&Prefer not to say                           & 1\\
\midrule
\multirow{3}{*}{\rotatebox[origin=c]{90}{\textbf{IT}}}                  

&No& 52\\
&Yes &45\\
&Prefer not to say& 3\\
\midrule

\multirow{6}{*}{\rotatebox[origin=c]{90}{\textbf{FB Usage}}}                  
&Every day&                 60\\
&A few times a week&        27\\
&About once a week      &    1\\
&A few times a month &       5\\
&Once a month       &        5\\
&Less than once a month&     2\\

\bottomrule
%\cmidrule(l){2-3} 
\end{tabular}
  \label{tab:demographics}
\end{table}